\documentclass[a4paper,UKenglish,cleveref, autoref, thm-restate]{lipics-v2021}

\hideLIPIcs
\usepackage[ruled, vlined, linesnumbered]{algorithm2e}
\usepackage{numprint}
\npdecimalsign{.} %
\usepackage{makeidx}
\usepackage[usenames,dvipsnames]{xcolor}
\usepackage{color}
\usepackage{ulem}
\usepackage{hyperref}
\usepackage{xspace}
\usepackage{todonotes}
\usepackage{booktabs}
\usepackage{mathtools}
\usepackage{csvsimple}
\usepackage{caption}[=v1]

\newcommand{\algoname}[1]{\textsc{#1}}

\newcommand{\ie}{i.e.,\xspace~}

\newcommand{\eg}{e.g.,\xspace~}

\newcommand{\etal}{et~al.}

\def\comment#1{}

\def\withcomments{
  \newcounter{mycommentcounter}
   \def\comment##1{\refstepcounter{mycommentcounter}%
    \ifhmode%
     \unskip%
     {\dimen1=\baselineskip \divide\dimen1 by 2 %
       \raise\dimen1\llap{\tiny\bfseries \textcolor{red}{-\themycommentcounter-}}}\fi%
     \marginpar[{\renewcommand{\baselinestretch}{0.8}%
       \hspace*{3em}\begin{minipage}{5em}\footnotesize [\themycommentcounter]: \raggedright ##1\end{minipage}}]{\renewcommand{\baselinestretch}{0.8}%
       \begin{minipage}{5em}\footnotesize [\themycommentcounter]: \raggedright ##1\end{minipage}}}
  }

\definecolor{darkgreen}{RGB}{0,200,100}
\definecolor{orange}{RGB}{255,80,0}

\newcommand{\aStop}{p}
\newcommand{\stopA}{p}
\newcommand{\stopB}{q}
\newcommand{\stopC}{r}
\newcommand{\aLine}{L}
\newcommand{\aTrip}{T}
\newcommand{\tripA}{T_a}
\newcommand{\tripB}{T_b}
\newcommand{\stopEvent}[2]{#1[#2]}
\newcommand{\tripSegment}[3]{#1[#2,#3]}
\newcommand{\lineSegment}[3]{#1[#2,#3]}
\newcommand{\transfer}[4]{\stopEvent{#1}{#2} \to \stopEvent{#3}{#4}}
\newcommand{\stops}{\mathcal{S}}
\newcommand{\lines}{\mathcal{L}}
\newcommand{\trips}{\mathcal{T}}
\newcommand{\footpaths}{\mathcal{F}}
\newcommand{\paretoset}{\mathcal{P}}
\newcommand{\transfers}{\mathfrak{T}}
\newcommand{\aTransfer}{\mathfrak{t}}
\newcommand{\successorTransfers}{\transfers^\uparrow}
\newcommand{\sourceStop}{\aStop_s}
\newcommand{\targetStop}{\aStop_t}
\newcommand{\departureTime}{\uptau_\mathrm{dep}}
\newcommand{\arrivalTime}{\uptau_\mathrm{arr}}
\newcommand{\graph}{G}
\newcommand{\vertices}{V}
\newcommand{\numVertices}{n}
\newcommand{\numEdges}{m}
\newcommand{\edges}{E}
\newcommand{\aVertex}{v}
\newcommand{\vertexA}{u}
\newcommand{\vertexB}{v}
\newcommand{\sourceVertex}{\aVertex_s}
\newcommand{\targetVertex}{\aVertex_t}
\newcommand{\anEdge}{e}
\newcommand{\layoutGraph}{\graph_L}
\newcommand{\layoutEdges}{\edges_L}
\newcommand{\layoutEdgeWeight}{\edgeWeight_L}
\newcommand{\aJourney}{J}
\newcommand{\anItinerary}{\mathcal{I}}
\newcommand{\edgeWeight}{c}
\newcommand{\aPath}{P}
\newcommand{\numCells}{k}
\newcommand{\partition}{r}
\newcommand{\aFlag}{b}
\newcommand{\aFixedFlag}{\hat{\aFlag}}
\newcommand{\reachedIndex}{R}

\newcommand{\connections}{X}
\newcommand{\tb}{\algoname{TB}\xspace}
\newcommand{\arcflagtb}{\algoname{Arc-Flag TB}\xspace}

\newcommand{\bufferedTrip}{\aTrip_\mathrm{min}}
\newcommand{\bufferedJourney}{\aJourney_\mathrm{min}}
\newcommand{\bufferedParetoSet}{\paretoset_\mathrm{buf}}

\withcomments

\makeatletter
\newcommand\footnoteref[1]{\protected@xdef\@thefnmark{\ref{#1}}\@footnotemark}
\makeatother

\usepackage{upgreek} %
\usepackage{siunitx}

\newcommand{\absoluteVal}[1]{\left\vert #1 \right\vert}

\newcommand{\arrtime}[2]{\uptau_{\mathrm{arr}}( \stopEvent{#1}{#2} )}
\newcommand{\deptime}[2]{\uptau_{\mathrm{dep}}( \stopEvent{#1}{#2} )}

\newcommand{\stopOfStopEvent}[2]{\aStop( \stopEvent{#1}{#2} )}
\newcommand{\transfertime}[2]{\Delta\uptau_{\mathrm{fp}}( #1, #2 )}

\newcommand{\codestyling}[1]{\texttt{#1}}
\newcommand{\printTime}[3]{\num[minimum-integer-digits = 2]{#1}{:}\num[minimum-integer-digits = 2]{#2}{:}\num[minimum-integer-digits = 2]{#3}}

\newcommand{\gs}{\hphantom{\tiny$\cdot$}}

\usepackage{xcolor}
\definecolor{KITgreen}          {rgb}{0,    0.588,0.509}
\definecolor{KITgreen70}        {rgb}{0.3,  0.711,0.656}
\definecolor{KITgreen50}        {rgb}{0.5,  0.794,0.754}
\definecolor{KITgreen30}        {rgb}{0.7,  0.876,0.852}
\definecolor{KITgreen15}        {rgb}{0.85, 0.938,0.926}
\definecolor{KITblue}           {rgb}{0.274,0.392,0.666}
\definecolor{KITblue70}         {rgb}{0.492,0.574,0.766}
\definecolor{KITblue50}         {rgb}{0.637,0.696,0.833}
\definecolor{KITblue30}         {rgb}{0.782,0.817,0.9}
\definecolor{KITblue15}         {rgb}{0.891,0.908,0.95}
\definecolor{KITpalegreen}      {rgb}{0.509,0.745,0.235}
\definecolor{KITpalegreen70}    {rgb}{0.656,0.821,0.464}
\definecolor{KITpalegreen50}    {rgb}{0.754,0.872,0.617}
\definecolor{KITpalegreen30}    {rgb}{0.852,0.923,0.77}
\definecolor{KITpalegreen15}    {rgb}{0.926,0.961,0.885}
\definecolor{KITyellow}         {rgb}{0.98, 0.901,0.078}
\definecolor{KITyellow70}       {rgb}{0.986,0.931,0.354}
\definecolor{KITyellow50}       {rgb}{0.99, 0.95, 0.539}
\definecolor{KITyellow30}       {rgb}{0.994,0.97, 0.723}
\definecolor{KITyellow15}       {rgb}{0.997,0.985,0.861}
\definecolor{KITorange}         {rgb}{0.862,0.627,0.117}
\definecolor{KITorange70}       {rgb}{0.903,0.739,0.382}
\definecolor{KITorange50}       {rgb}{0.931,0.813,0.558}
\definecolor{KITorange30}       {rgb}{0.958,0.888,0.735}
\definecolor{KITorange15}       {rgb}{0.979,0.944,0.867}
\definecolor{KITbrown}          {rgb}{0.627,0.509,0.196}
\definecolor{KITbrown70}        {rgb}{0.739,0.656,0.437}
\definecolor{KITbrown50}        {rgb}{0.813,0.754,0.598}
\definecolor{KITbrown30}        {rgb}{0.888,0.852,0.758}
\definecolor{KITbrown15}        {rgb}{0.944,0.926,0.879}
\definecolor{KITred}            {rgb}{0.627,0.117,0.156}
\definecolor{KITred70}          {rgb}{0.739,0.382,0.409}
\definecolor{KITred50}          {rgb}{0.813,0.558,0.578}
\definecolor{KITred30}          {rgb}{0.888,0.735,0.747}
\definecolor{KITred15}          {rgb}{0.944,0.867,0.873}
\definecolor{KITlilac}          {rgb}{0.627,0,    0.47}
\definecolor{KITlilac70}        {rgb}{0.739,0.3,  0.629}
\definecolor{KITlilac50}        {rgb}{0.813,0.5,  0.735}
\definecolor{KITlilac30}        {rgb}{0.888,0.7,  0.841}
\definecolor{KITlilac15}        {rgb}{0.944,0.85, 0.92}
\definecolor{KITcyanblue}       {rgb}{0.313,0.666,0.901}
\definecolor{KITcyanblue70}     {rgb}{0.519,0.766,0.931}
\definecolor{KITcyanblue50}     {rgb}{0.656,0.833,0.95}
\definecolor{KITcyanblue30}     {rgb}{0.794,0.9,  0.97}
\definecolor{KITcyanblue15}     {rgb}{0.897,0.95, 0.985}
\definecolor{KITseablue}        {rgb}{0.196,0.313,0.549}
\definecolor{KITseablue70}      {rgb}{0.437,0.519,0.684}
\definecolor{KITseablue50}      {rgb}{0.598,0.656,0.774}
\definecolor{KITseablue30}      {rgb}{0.758,0.794,0.864}
\definecolor{KITseablue15}      {rgb}{0.879,0.897,0.932}
\definecolor{KITblack}          {rgb}{0,    0,    0}
\definecolor{KITblack90}        {rgb}{0.1,  0.1,  0.1}
\definecolor{KITblack80}        {rgb}{0.2,  0.2,  0.3}
\definecolor{KITblack75}        {rgb}{0.25, 0.25, 0.25}
\definecolor{KITblack70}        {rgb}{0.3,  0.3,  0.3}
\definecolor{KITblack60}        {rgb}{0.4,  0.4,  0.4}
\definecolor{KITblack50}        {rgb}{0.5,  0.5,  0.5}
\definecolor{KITblack40}        {rgb}{0.6,  0.6,  0.6}
\definecolor{KITblack30}        {rgb}{0.7,  0.7,  0.7}
\definecolor{KITblack25}        {rgb}{0.75, 0.75, 0.75}
\definecolor{KITblack20}        {rgb}{0.8,  0.8,  0.8}
\definecolor{KITblack10}        {rgb}{0.9,  0.9,  0.9}
\definecolor{KITwhite}          {rgb}{1,    1,    1}

\colorlet{nodeColor}{black!80}
\colorlet{edgeColor}{black!100}

\usepackage{quiver}
\usepackage{tikz}
\usepackage{pgfplots}
\usepgfplotslibrary{groupplots}
\usetikzlibrary{positioning,shapes,arrows,fit}
\pgfplotsset{compat=newest}
\pgfkeys{/pgf/number format/.cd, set thousands separator={\,}}

\tikzstyle{vertex}=[circle,line width=.5pt,minimum size=0.1pt]
\tikzstyle{routeArrow}=[->, >=stealth]
\tikzstyle{edgeArrow}=[->, >=stealth]

\usepackage{dsfont} %
\usepackage{makecell} 	%
\hideLIPIcs

\title{Arc-Flags Meet Trip-Based Public Transit Routing}

\author{Ernestine Großmann}{Heidelberg University, Heidelberg, Germany}{e.grossmann@informatik.uni-heidelberg.de}{https://orcid.org/0000-0002-9678-0253}{}
\author{Jonas Sauer}{Karlsruhe Institute of Technology, Karlsruhe, Germany}{jonas.sauer2@kit.edu}{https://orcid.org/0000-0002-7196-7468}{}
\author{Christian Schulz}{Heidelberg University, Heidelberg, Germany}{christian.schulz@informatik.uni-heidelberg.de}{https://orcid.org/0000-0002-2823-3506}{}
\author{Patrick Steil}{Heidelberg University, Heidelberg, Germany}{patrick.steil@informatik.uni-heidelberg.de}{https://orcid.org/0000-0003-3282-4533}{}

\authorrunning{E. Großmann \etal} %

\Copyright{Ernestine Großmann, Jonas Sauer, Christian Schulz, Patrick Steil} %

\keywords{Public transit routing, graph algorithms, algorithm engineering} %

\category{} %

\relatedversion{} %

\funding{We acknowledge support by DFG grants SCHU 2567/3-1 and WA 654/23-2.}
 
\acknowledgements{We want to thank Dr.~Patrick Brosi for providing us with the Germany dataset and Sascha Witt for providing us with the source code for \algoname{TB-CST}.}%

\begin{document}

\maketitle

\begin{abstract}
We present \arcflagtb, a journey planning algorithm for public transit networks which combines \algoname{Trip-Based Public Transit Routing (\tb)}~\cite{Wit15} with the \algoname{Arc-Flags}~\cite{Moe06} speedup technique.
Compared to previous attempts to apply \algoname{Arc-Flags} to public transit networks, which saw limited success, our approach uses stronger pruning rules to reduce the search space.
Our experiments show that \arcflagtb achieves a speedup of up to two orders of magnitude over \tb, offering query times of less than a millisecond even on large countrywide networks.
Compared to the state-of-the-art speedup technique \algoname{Trip-Based Public Transit Routing Using Condensed Search Trees} (\algoname{TB-CST})~\cite{Wit16}, our algorithm achieves similar query times but requires significantly less additional memory.
Other state-of-the-art algorithms which achieve even faster query times, \eg \algoname{Public Transit Labeling}~\cite{Del15}, require enormous memory usage.
In contrast, \arcflagtb offers a tradeoff between query performance and memory usage due to the fact that the number of regions in the network partition required by our algorithm is a configurable parameter.
We also identify an issue in the transfer precomputation of \tb that affects both \algoname{TB-CST} and \arcflagtb, leading to incorrect answers for some queries.
This has not been previously recognized by the author of \algoname{TB-CST}.
We provide discussion on how to resolve this issue in the future.
Currently, \arcflagtb answers 1--6\% of queries incorrectly, compared to over 20\% for \algoname{TB-CST} on some networks.

\end{abstract}
\section{Introduction}
\label{sec:introduction}
Interactive journey planning applications have become a part of our everyday lives.
To provide routing information in real time, these applications rely on very fast query algorithms.
While \algoname{Dijkstra's Algorithm}~\cite{Dij59} solves the shortest path problem in quasi-linear time, it still takes several seconds on continental-sized networks, which is too slow for interactive use.
Practical applications therefore rely on \emph{speedup techniques}, which compute auxiliary data in a preprocessing phase and then use this data to speed up the query phase.
Recent decades have seen the development of many successful speedup techniques for route planning on road networks~\cite{Bas16b}.
These achieve query times of less than a millisecond with only moderate preprocessing time and space consumption.

For public transit networks, the state of the art is not as satisfactory.
In order to achieve query times below a millisecond on large country-sized networks, existing techniques must precompute data in the tens to hundreds of gigabytes.
This discrepancy has been explained by the fact that road networks exhibit beneficial structural properties which are not as pronounced in public transit networks~\cite{Bas09}.
An additional challenge is that passengers in public transportation systems typically consider more criteria than just the travel time when evaluating journeys.
Most recent algorithms in the literature Pareto-optimize at least two criteria: arrival time and the number of used trips.
For these reasons, a speedup technique which achieves very low query times with only a moderate amount of precomputed data has remained elusive. 

\subparagraph*{State of the Art.}
\label{par:stateoftheart}
For this work, we consider algorithms for journey planning in public transit networks which Pareto-optimize arrival time and number of trips. 
For a more general overview of journey planning algorithms, we refer to~\cite{Bas16b}.
The classical approach is to model the public transit timetable as a graph and then apply a multicriteria variant of \algoname{Dijkstra's Algorithm}~\cite{Han80,Mue07b,Dis08}.
The time-dependent and time-expanded approaches are the two most prominent ways of modelling the timetable.
In the~\emph{time-dependent} model~\cite{Bro04,Pyr08}, stops in the network are represented by nodes in the graph and connections between them as edges with a time-dependent, piecewise linear travel time function.
This yields a compact graph but requires a time-dependent version of \algoname{Dijkstra's Algorithm}.
By contrast, the~\emph{time-expanded} model~\cite{Mue07b,Pyr08} introduces a node for each event in the timetable (\eg a vehicle arriving or departing from a stop).
Edges connect consecutive events of the same trip and events between which a transfer is possible.
The resulting graph is significantly larger but has scalar edge weights, allowing \algoname{Dijkstra's Algorithm} to be applied without modification. 

Using a graph-based model has the advantage that speedup techniques for \algoname{Dijkstra's Algorithm} can be applied.
However, the achieved speedups are much smaller than on road networks~\cite{Bas09,Bau11}.
A notable technique which has been applied to bicriteria optimization in public transit networks is \algoname{Arc-Flags}~\cite{Moe06}.
Its basic idea is to partition the graph into regions and to compute a \emph{flag} for each combination of edge and region, which indicates whether the edge is required to reach the region.
\algoname{Dijkstra's Algorithm} can then be sped up by ignoring unflagged edges.
\algoname{Arc-Flags} has been applied to both time-dependent~\cite{Ber09} and time-expanded~\cite{Del09c} graphs, although only arrival time was optimized in the latter case.
This yielded speedups of~3 and~4, respectively, whereas \algoname{Arc-Flags} on road networks achieves speedups of over~500~\cite{Moe06}.

More recent algorithms do not model the timetable as a graph but employ more cache-efficient data structures to achieve faster query times.
Notable examples are \algoname{RAPTOR}~\cite{Del15b} and~\algoname{Trip-Based Public Transit Routing}~(TB)~\cite{Wit15}.
The latter employs a lightweight preprocessing phase which precomputes relevant transfers between individual trips.
This yields query times in the tens of milliseconds even on large networks, which is a significant improvement over graph-based techniques.

Algorithms which reduce query times to the sub-millisecond range do so by precomputing auxiliary data whose size is quadratic in the size of the network.
\algoname{Public Transit Labeling}~(\algoname{PTL})~\cite{Del15} adapts the ideas of \algoname{Hub Labeling}~\cite{Coh03} to time-expanded graphs.
While this yields query times of a few microseconds, it requires tens of gigabytes of space on metropolitan networks.
Moreover, this does not include the additional overhead required for~\emph{journey unpacking}, \ie retrieving descriptions of the optimal journeys, which would increase the size of the auxiliary data into the hundreds of gigabytes.
\algoname{Transfer Patterns}~(\algoname{TP})~\cite{Bas10} employs a preprocessing phase which essentially answers every possible query in advance.
Since storing a full description of every optimal journey would require too much space, TP condenses this information into a generalized search graph for each possible source stop, which is then explored during the query phase.
On the network of Germany, TP answers queries in less than a millisecond but requires hundred of hours of preprocessing time and over 100\,GB of space.
\algoname{Scalable Transfer Patterns}~\cite{Bas16} reduces the preprocessing effort with a clustering-based approach, but the resulting query times are only barely competitive with TB.
\algoname{Trip-Based Routing Using Condensed Search Trees}~(\algoname{TB-CST})~\cite{Wit16} re-engineers the ideas of TP with a faster, \tb-based preprocessing algorithm and by splitting the computed search graphs in order to save space.
However, there is an issue in the transfer precomputation of \tb that leads to incorrect answers for some queries.
This has not been previously recognized by the author of \algoname{TB-CST}.

\subparagraph*{Contribution.}
We revisit the concept of \algoname{Arc-Flags} for public transit journey planning.
In contrast to previous approaches, we use modern \tb-based algorithms in preprocessing and query phases.
The high cache efficiency and stronger pruning rules of these algorithms drastically reduce the search space and running times. 
The resulting algorithm, \arcflagtb, matches or exceeds the performance of \algoname{TB-CST} with a similar precomputation time and significantly lower space consumption.
Compared to \tb, it achieves a speedup of one order of magnitude on metropolitan networks and two orders of magnitude on country networks.
Since the number of regions in the underlying network partition is a configurable parameter, \arcflagtb additionally offers a tradeoff between query performance and the size of the precomputed data.

We identify an issue in the transfer precomputation of TB, which both \algoname{TB-CST} and \arcflagtb rely on.
As a result, both algorithms answer some queries incorrectly.
We discuss how this issue can be resolved in the future.
In its current configuration, \arcflagtb answers~$1$--$6\%$ of queries incorrectly, depending on the network, compared to over~$20\%$ for \algoname{TB-CST} on some networks.
Altogether, we show that \algoname{Arc-Flags} for public transit networks has more potential than previously thought.

\section{Preliminaries}
\label{sec:preliminaries}

\subsection{Basic Concepts}
\subparagraph*{Public Transit Network.}
A~\emph{public transit network} is a 4-tuple~$(\stops,\lines,\trips,\footpaths)$ consisting of a set of stops~$\stops$, a set of lines~$\lines$, a set of trips $\trips$, and a set of footpaths~$\footpaths\subseteq\stops\times\stops$.
A \textit{stop}~$\aStop\in\stops$ is a location where a vehicle stops and passengers can enter or exit the vehicle.
A \textit{trip} is a sequence~$\aTrip=\left< \stopEvent{\aTrip}{0}, \stopEvent{\aTrip}{1}, \dots\right>$ of \textit{stop events}, where each stop event~$\stopEvent{\aTrip}{i}$ has an associated arrival time~$\arrtime{\aTrip}{i}$, departure time~$\deptime{\aTrip}{i}$ and stop~$\stopOfStopEvent{\aTrip}{i} \in \stops$.
We denote the number of stop events in~$\aTrip$ as~$\absoluteVal{\aTrip}$.
Trips with the same stop sequence that do not overtake each other are grouped into \emph{lines}.
A trip~$\tripA \in \trips$ overtakes another trip~$\tripB \in \trips$ if there are stops~$\stopA, \stopB \in \stops$ such that~$\tripB$ arrives (or departs) later at~$\stopA$ than~$\tripA$, but~$\tripA$ arrives (or departs) earlier than~$\tripB$ at~$\stopB$.
The set of all trips belonging to a line~$\aLine$ is denoted as~$\trips(\aLine)$.
Since trips $\tripA, \tripB \in \trips(\aLine)$ cannot overtake each other, we can define a total ordering
\begin{align*}
	\tripA \preceq \tripB &\iff \forall i \in \left[ 0, \absoluteVal{\tripA}\right): \arrtime{\tripA}{i} \leq \arrtime{\tripB}{i}\\
	\tripA \prec \tripB &\iff \tripA \preceq \tripB \wedge \exists i \in \left[ 0, \absoluteVal{\tripA}\right): \arrtime{\tripA}{i} < \arrtime{\tripB}{i}.
\end{align*}
A~\emph{footpath}~$(\stopA,\stopB)\in\footpaths$ allows passengers to transfer between stops~$\stopA$ and~$\stopB$ with the~\emph{transfer time}~$\transfertime{\stopA}{\stopB}$.
If no footpath between~$\stopA$ and~$\stopB$ exists, we define~$\transfertime{\stopA}{\stopB} = \infty$.
If $\stopA = \stopB$, then $\transfertime{\stopA}{\stopB} = 0$.
We require that the set of footpaths is transitively closed and fulfills the triangle inequality, \ie if there are stops~$\stopA,\stopB,\stopC\in\stops$ with~$(\stopA,\stopB)\in\footpaths$ and~$(\stopB,\stopC)\in\footpaths$, then there must be a footpath~$(\stopA,\stopC)\in\footpaths$ with~$\transfertime{\stopA}{\stopC}\leq\transfertime{\stopA}{\stopB}+\transfertime{\stopB}{\stopC}$.

A~\textit{trip segment}~$\tripSegment{\aTrip}{i}{j}$ ($0 \leq i < j < |\aTrip|$) is the subsequence of trip~$\aTrip$ between the two stop events~$\stopEvent{\aTrip}{i}$ and~$\stopEvent{\aTrip}{j}$.
A \textit{transfer}~$\transfer{\tripA}{i}{\tripB}{j}$ represents a passenger transferring from~$\tripA$ to~$\tripB$ at the corresponding stop events.
Note that this requires~$\arrtime{\tripA}{i}+\transfertime{\stopOfStopEvent{\tripA}{i}}{\stopOfStopEvent{\tripB}{j}}\leq\deptime{\tripB}{j}$.
A \textit{journey}~$\aJourney$ from a source stop~$\sourceStop$ to a target stop~$\targetStop$ is an sequence of trip segments such that every pair of consecutive trip segments is connected by a transfer.
In addition, a journey contains an \textit{initial} and \textit{final footpath}, where the initial footpath connects~$\sourceStop$ to the first stop event, and the final footpath connects the last stop event to~$\targetStop$.

\subparagraph*{Problem Statement.}
A journey~$\aJourney$ from~$\sourceStop$ to~$\targetStop$ is evaluated according to two criteria: its arrival time at~$\targetStop$, and the number of trips used by~$\aJourney$.
We say that~$\aJourney$ \textit{weakly dominates} another journey~$\aJourney'$ if~$\aJourney$ is not worse than~$\aJourney'$ in either of the two criteria.
Moreover, $\aJourney$ \textit{strongly dominates}~$\aJourney'$ if~$\aJourney$ weakly dominates~$\aJourney'$ and~$\aJourney$ is strictly better in at least one criterion.
Given source and target stops~$\sourceStop,\targetStop\in\stops$ and an earliest departure time~$\departureTime$ at~$\sourceStop$, a journey~$\aJourney$ from~$\sourceStop$ to~$\targetStop$ is \emph{feasible} if its departure time at~$\sourceStop$ is not earlier than~$\departureTime$.
A~\emph{Pareto set}~$\paretoset$ is a set of journeys such that~$\paretoset$ has minimal size and every feasible journey is weakly dominated by a journey in~$\paretoset$.
Given source and target stops~$\sourceStop,\targetStop\in\stops$ and a departure time~$\departureTime$, the \textit{fixed departure time problem} asks for a Pareto set with respect to the two criteria arrival time and number of trips.
For the \textit{profile problem}, we are given an interval $\left[\uptau_1, \uptau_2\right]$ of possible departure times in addition to~$\sourceStop$ and~$\targetStop$.
Here, the objective is to find the union of the Pareto sets for each distinct departure time $\uptau \in \left[ \uptau_1, \uptau_2\right]$.
In the~\emph{full-range profile problem}, the departure time interval spans the entire service duration of the network.

\subparagraph*{Graph.}
A directed, weighted graph~$\graph = (\vertices, \edges, \edgeWeight)$ is a triple consisting of a set of nodes~$\vertices$, a set of edges~$\edges \subseteq \vertices \times \vertices$, and an edge weight function~$\edgeWeight: \edges \to \mathbb{R}$.
We denote by~$\numVertices = \absoluteVal{\vertices}$ the number of nodes and~$\numEdges = \absoluteVal{\edges}$ the number of edges.
A \textit{path}~$\aPath = \left< \aVertex_1, \aVertex_2, \dots, \aVertex_k\right>$ is a sequence of nodes between~$\aVertex_1$ and~$\aVertex_k$ such that each pair of consecutive nodes is connected by an edge.
The weight of a path is the sum of the weights of all edges in the path. 
A path~$\aPath = \left< \sourceVertex, \dots, \targetVertex \right>$ between a source node~$\sourceVertex$ and a target node~$\targetVertex$ is called the \textit{shortest path} if there is no other path between~$\sourceVertex$ and~$\targetVertex$ with a smaller weight.

Given a value~$\numCells\in\mathbb{N}$ and a graph~$\graph = \left(\vertices, \edges, \edgeWeight\right)$, a ($\numCells$-way) \emph{partition} of~$\graph$ is a function~$\partition : \vertices \to \left\{1, \dots, \numCells\right\}$ which partitions the node set~$\vertices$ into~$\numCells$ \textit{cells}.
The set of nodes in cell~$i$ is denoted as~$\vertices_i \coloneqq \partition^{-1}(i)$.
An edge~$(\vertexA,\vertexB)$ is called a~\emph{cut edge} if its endpoints~$\vertexA$ and~$\vertexB$ belong to different cells.
A node is called a~\emph{boundary node} if it is incident to a cut edge.
The partition is called~\emph{balanced} for an imbalance parameter~$\varepsilon > 0$ if the size of each cell~$\vertices_i$ is bounded by

\begin{equation*}
	\absoluteVal{\vertices_i} \leq \left( 1 + \varepsilon \right) \left\lceil \frac{\absoluteVal{\vertices}}{\numCells}\right\rceil.
\end{equation*}
The \emph{graph partitioning problem} asks for a balanced partition that minimizes the weighted sum of all~cut~edges.

\subsection{Related Work}
\subparagraph*{Trip-Based Public Transit Routing.}
\label{par:tb}
The \algoname{Trip-Based Public Transit Routing (TB)} algorithm~\cite{Wit15} solves the fixed departure time problem on a public transit network.
It employs a precomputation phase, which first generates all possible transfers between stop events.
Then, using a set of pruning rules, transfers which are not required to answer queries are discarded.
We denote the remaining set of transfers as~$\transfers$.
Note that~$\transfers$ may still contain transfers which do not occur in any Pareto-optimal journey.

The \tb query algorithm is a modified breadth-first search on the set of trips and the precomputed transfers.
The algorithm tracks which parts of the network have already been explored by maintaining a~\emph{reached index}~$\reachedIndex(\aTrip)$ for each trip~$\aTrip$.
This is the index of the first reached stop event of~$\aTrip$, or~$\absoluteVal{\aTrip}$ if none have been reached yet.
The \tb query operates in~\emph{rounds}, where round~$i$ finds Pareto-optimal journeys which use~$i$ trips.
Each round maintains a FIFO (first-in-first-out) queue of newly reached trip segments; these are then scanned during the round.
A trip segment~$\tripSegment{\tripA}{i}{j}$ is scanned by iterating over the stop events~$\stopEvent{\tripA}{k}$ with~$i\leq{}k\leq{}j$ and relaxing all outgoing transfers~$(\stopEvent{\tripA}{k},\stopEvent{\tripB}{\ell})\in\transfers$.
If~$\ell<\reachedIndex(\tripB)$, then the trip segment~$\tripSegment{\tripB}{\ell}{\reachedIndex(\tripB)-1}$ is added to the queue for the next round.
Additionally, for every succeeding trip~$\tripB'$ of the same line with~$\tripB \preceq \tripB'$, the reached index~$\reachedIndex(\tripB')$ is set to $\min\left(\reachedIndex(\tripB'), \ell\right)$.
This ensures that the search only enters the earliest reachable trip of each line, a principle we call \emph{line pruning}.

Profile-\tb is an extension of \tb which solves the profile problem.
It exploits the observation that journeys with a later departure time weakly dominate journeys with an earlier arrival time if they are equivalent or better in the other criteria.
Therefore, it collects all possible departure times at~$\sourceStop$ that lie within the departure time interval~$[\uptau_1,\uptau_2]$ and processes them in descending order.
For each departure time, a run of the \tb query algorithm is performed.
All data structures, including reached indices, are not reset between runs.
This allows results from earlier runs to prune suboptimal results in the current run, a principle called \emph{self-pruning}.
In order to obtain correct results, the definition of reached indices must be modified slightly.
For each trip~$\aTrip$ and each possible number of trips~$i$, the algorithm now maintains a reached index~$\reachedIndex_i(\aTrip)$, which is the index of the first stop event in~$\aTrip$ which was reached with~$i$ or fewer trips.
Whenever~$\reachedIndex_i(\aTrip)$ is updated to~$\min(\reachedIndex_i(\aTrip),k)$ for some value~$k$, the same is done for the reached indices~$\reachedIndex_j(\aTrip)$ with~$j \geq i$.

\subparagraph*{Condensed Search Trees.}
\label{par:tb-cst}
\algoname{Trip-Based Routing Using Condensed Search Trees}~(\algoname{TB-CST})~\cite{Wit16} employs Profile-\tb to precompute search graphs which allow for extremely fast queries.
The preprocessing phase solves the full-range profile problem for every possible pair of source and target stops by running a modified \emph{one-to-all} version of Profile-\tb from every stop.
Consider the Profile-\tb search for a source stop~$\sourceStop$.
After each \tb run, all newly found Pareto-optimal journeys are unpacked.
This yields a breadth-first search tree with~$\sourceStop$ as the root, trip segments as inner nodes, the reached target stops as leaves, and footpaths and transfers as edges.
The search trees of all runs are merged into the \emph{prefix tree} of~$\sourceStop$.
Here, each trip segment~$\tripSegment{\aTrip}{i}{j}$ is replaced with a tuple~$(\aLine,i)$ consisting of the line~$\aLine$ with~$\aTrip\in\trips(\aLine)$ and the stop index~$i$ where the line is entered.
	
To answer one-to-all queries, Profile-\tb additionally maintains an earliest arrival time~$\arrivalTime(\aStop,n)$ for each stop~$\aStop$ and number of trips~$n$.
Like the reached indices, these arrival times are not reset between runs.
When scanning a stop event~$\stopEvent{\aTrip}{k}$ in round~$n$, the algorithm iterates over all stops~$\aStop$~with~$\transfertime{\stopOfStopEvent{\aTrip}{k}}{\aStop}<\infty$ and computes~$\overline{\arrivalTime}=\arrtime{\aTrip}{k}+\transfertime{\stopOfStopEvent{\aTrip}{k}}{\aStop}$.
If~$\overline{\arrivalTime}<\arrivalTime(\aStop,n)$, then the best known journey to~$\aStop$ with~$n$ trips was improved, so~$\arrivalTime(\aStop,m)$ is set to~$\min(\overline{\arrivalTime},\arrivalTime(\aStop,m))$ for all~$m\geq{}n$.

To answer a query between source stop~$\sourceStop$ and target stop~$\targetStop$, \algoname{TB-CST} constructs a~\emph{query graph} from the prefix tree of~$\sourceStop$ by extracting all paths which lead to a leaf representing~$\targetStop$.
Then a variant of \algoname{Dijkstra's Algorithm} is run on the query graph.
Since the prefix tree only provides information about lines, but not specific trips, these must be reconstructed during the query.
When relaxing an edge from~$\sourceStop$ to the first used line, the earliest reachable trip is identified based on the departure time at~$\sourceStop$.
When relaxing an edge between lines~$\aLine_1$ and~$\aLine_2$, the used trip~$\aTrip_1$ of~$\aLine_1$ is already known, so the algorithm explores the outgoing transfers of~$\aTrip_1$ in~$\transfers$ to find the earliest reachable trip~$\aTrip_2$ of~$\aLine_2$.

The space required to store all prefix trees can be reduced by extracting \emph{postfix trees}.
Consider the prefix tree for a source stop~$\sourceStop$.
For each path from the root to a leaf representing a target stop~$\targetStop$, a~\emph{cut node} is chosen.
The subpath from the cut node to the leaf is then removed from the prefix tree of~$\sourceStop$ and added to the postfix tree of~$\targetStop$.
Since many of these extracted subpaths occur in multiple prefix trees, moving them into a shared postfix tree considerably reduces the memory consumption.
To construct the query graph for a source stop~$\sourceStop$ and target stop~$\targetStop$, the prefix tree~$\sourceStop$ and the postfix tree of~$\targetStop$ are spliced back together at the cut nodes.

\subparagraph*{Arc-Flags.}
\algoname{Arc-Flags} is a speedup technique for \algoname{Dijkstra's Algorithm}.
Its basic idea is to precompute \emph{flags} for each edge, which indicate whether the edge is necessary to reach a certain region of the graph.
This allows \algoname{Dijkstra's Algorithm} to reduce the search space during a query by ignoring edges which are not flagged for the target region.

Given a weighted graph~$\graph=\left(\vertices, \edges, \edgeWeight\right)$, the preprocessing phase of \algoname{Arc-Flags} performs two steps:
First, it computes a partition~$\partition: \vertices \to \{1,\dots,\numCells\}$ of the node set into~$\numCells$ cells, where~$\numCells$ is a freely chosen parameter.
Then, a~\emph{flags function}~$\aFlag : \edges\times\left\{1, \dots, \numCells\right\} \to \left\{0, 1\right\}$ is computed.
Each individual value~$\aFlag(\anEdge, i)$ for an edge~$\anEdge$ and a cell~$i$ is called a \emph{flag}, hence the name \algoname{Arc-Flags}.
The flags function must have the following property: for each pair of source node~$\sourceVertex$ and target node~$\targetVertex$, there is at least one shortest path~$\aPath$ from~$\sourceVertex$ to~$\targetVertex$ such that~$\aFlag(\anEdge,\partition(\targetVertex))=1$ for every edge~$\anEdge$ in~$\aPath$.
With this precomputed information, a shortest path query between~$\sourceVertex$ and~$\targetVertex$ can be answered by running \algoname{Dijkstra's Algorithm} but only relaxing edges~$\anEdge$ for which~$\aFlag(\anEdge,\partition(\targetVertex))=1$.
The parameter~$\numCells$ imposes a tradeoff between query speed and memory consumption.
The space required to store the flags is in~$\Theta\left(\numCells\numEdges\right)$, which is manageable for~$\numCells \ll \numVertices$.
On the other hand, the search space of the query decreases for larger values of~$\numCells$, since fewer flags will be set to~$1$ if the target cell is smaller.

Flags can be computed naively by solving the \textit{all-pairs shortest path} problem, \ie computing the shortest path between every pair of nodes.
However, this requires~$\Omega(\numVertices^2)$ precomputation time.
The precomputation can be sped up exploiting the observation that every shortest path that leads into a cell must pass through a boundary node.
Thus, it is sufficient to compute backward shortest-path trees from all boundary nodes.
For more details, see~\cite{Hil09}.

\subparagraph*{Arc-Flags for Public Transit Networks.}
Berger \etal~\cite{Ber09} applied \algoname{Arc-Flags} to a time-dependent graph model in a problem setting which asks for \emph{all} Pareto-optimal paths, including duplicates (\ie Pareto-optimal paths which are equivalent in both criteria).
They observed that for nearly every combination of edge~$\anEdge$ and cell~$i$, there is at least one point in time during which~$\anEdge$ occurs on a Pareto-optimal path to a node in cell~$i$.
To solve this problem, the authors divided the service period of the network into two-hour intervals and computed a flag for each combination of edge, cell and time interval.
However, this approach merely achieved a speedup of~$\approx 3$ over \algoname{Dijkstra's Algorithm}.

Time resolution is not an issue in time-expanded graphs, where each node is associated with a specific point in time.
However, Delling \etal~\cite{Del09c} observed a different problem when applying \algoname{Arc-Flags} to a time-expanded graph, even when optimizing only arrival time.
Since the arrival time of a path depends only on its target node, all valid paths are optimal.
Delling \etal~therefore evaluated various tiebreaking strategies to decide which optimal paths should be flagged.
The most successful strategy only achieved a speedup of~$\approx 4$ over \algoname{Dijkstra's Algorithm}.
In the same paper, Delling \etal~proposed a pruning technique called \emph{Node-Blocking}, which applies the principle of line pruning to \algoname{Dijkstra's Algorithm} in time-expanded graphs.
The authors observed that Node-Blocking conflicts with their tiebreaking choices for \algoname{Arc-Flags}, leading to incorrectly answered queries.
Therefore, they only evaluated \algoname{Arc-Flags} without Node-Blocking.

\section{\arcflagtb}
\label{sec:arcflagtbalgorithm}

We now present the core ideas of our new algorithm \arcflagtb, which applies the main idea of \algoname{Arc-Flags} to \tb.
We first explain the general idea and then discuss details and optimizations.
Finally, we compare our approach to similar algorithms.

The \arcflagtb precomputation performs two tasks:
First it partitions the set~$\stops$ of stops into~$\numCells$ cells, which yields a partition function~$\partition:\stops \to \left\{1, \dots, \numCells\right\}$.
Then it computes a flag for each transfer~$\aTransfer\in\transfers$ and cell~$i$ which indicates whether~$\aTransfer$ is required to reach any target stops in cell~$i$.
Formally, this yields a flags function~$\aFlag : \transfers \times \left\{1, \dots, \numCells\right\} \to \left\{0, 1\right\}$ with the following property: for each query with source stop~$\sourceStop$, target stop~$\targetStop$ and departure time~$\departureTime$, there is a Pareto set~$\paretoset$ such that~$\aFlag(\aTransfer,\partition(\targetStop))=1$ for every transfer~$\aTransfer=\transfer{\tripA}{i}{\tripB}{j}$ which occurs in a journey~$\aJourney\in\paretoset$.
A query between source stop~$\sourceStop$ and target stop~$\targetStop$ is answered by running the \tb query algorithm with one modification: a transfer~$\aTransfer\in\transfers$ is only explored if the flag for the target cell is set to~$1$, \ie $\aFlag(\aTransfer,\partition(\targetStop))=1$.

\subsection{Partitioning}
\label{subsec:Partitioning}

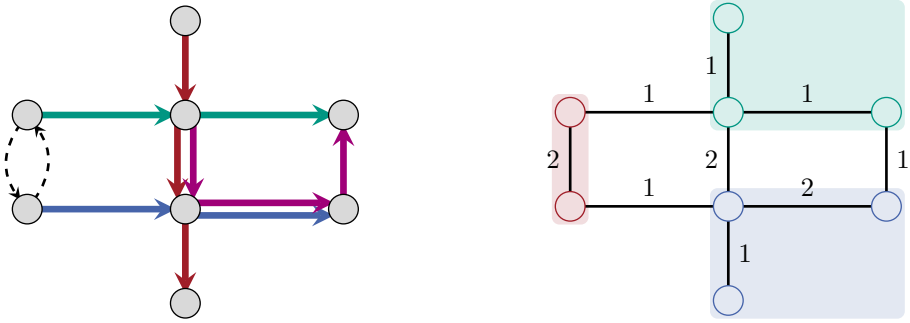
\begin{figure}[tp!]
    \caption{\textit{Left:} An example network with stops and nodes, trips as colored solid edges and transfers as dashed edges.
        \textit{Right:} The corresponding layout graph with edges weighted according to the number of connections they represent.
        Nodes are grouped according to a possible $3$-way partition of the graph.
    }
    \label{fig:metisgraph}
    \centering
    \begin{tikzpicture}[scale=1.04]
\begin{scope}[yscale=0.8]
    \node (v1) at (0.00, 0.00) {};%
    \node (v2) at (2.00, 0.00) {};%
    \node (v2left) at (1.90, 0.00) {};%
    \node (v2right) at (2.10, 0.00) {};%
    \node (v3) at (4.00, 0.00) {};%
    \node (v4) at (2.00, 1.50) {};%
    \node (v5) at (2.00, -1.50) {};%
    \node (v5left) at (1.90, -1.50) {};%
    \node (v5right) at (2.10, -1.50) {};%
    \node (v5up) at (2.00, -1.40) {};%
    \node (v5down) at (2.00, -1.60) {};%
    \node (v6) at (2.00, -3.00) {};%
    \node (v7) at (0.00, -1.50) {};%
    \node (v8) at (4.00, -1.50) {};%
    \node (v8up) at (4.00, -1.40) {};%
    \node (v8down) at (4.00, -1.60) {};%
    
    \draw [KITgreen, line width=2.5pt, routeArrow] (v1) -- (v2);
    \draw [KITgreen, line width=2.5pt, routeArrow] (v2) -- (v3);
    \draw [KITblue, line width=2.5pt, routeArrow] (v7) -- (v5);
    \draw [KITblue, line width=2.5pt, routeArrow] (v5down) -- (v8down);
    \draw [KITred, line width=2.5pt, routeArrow] (v4) -- (v2);
    \draw [KITred, line width=2.5pt, routeArrow] (v2left) -- (v5left);
    \draw [KITred, line width=2.5pt, routeArrow] (v5) -- (v6);
    \draw [KITlilac, line width=2.5pt, routeArrow] (v2right) -- (v5right);
    \draw [KITlilac, line width=2.5pt, routeArrow] (v5up) -- (v8up);
    \draw [KITlilac, line width=2.5pt, routeArrow] (v8) -- (v3);

    \draw [edgeColor, line width=1pt, dashed, edgeArrow]  (v1) to [bend right] (v7);
    \draw [edgeColor, line width=1pt, dashed, edgeArrow]  (v7) to [bend right] (v1);
    
    \node (v1v) at (v1) [vertex,draw=KITblack,fill=KITblack!15] {\gs};%
    \node (v2v) at (v2) [vertex,draw=KITblack,fill=KITblack!15] {\gs};%
    \node (v3v) at (v3) [vertex,draw=KITblack,fill=KITblack!15] {\gs};%
    \node (v4v) at (v4) [vertex,draw=KITblack,fill=KITblack!15] {\gs};%
    \node (v5v) at (v5) [vertex,draw=KITblack,fill=KITblack!15] {\gs};%
    \node (v6v) at (v6) [vertex,draw=KITblack,fill=KITblack!15] {\gs};%
    \node (v7v) at (v7) [vertex,draw=KITblack,fill=KITblack!15] {\gs};%
    \node (v8v) at (v8) [vertex,draw=KITblack,fill=KITblack!15] {\gs};%
\end{scope}
\end{tikzpicture}
    \hspace{2cm}
    \begin{tikzpicture}[scale=1.04]
\begin{scope}[yscale=0.8]
    \node (v1) at (0.00, 0.00) {};%
    \node (v2) at (2.00, 0.00) {};%
    \node (v3) at (4.00, 0.00) {};%
    \node (v4) at (2.00, 1.50) {};%
    \node (v5) at (2.00, -1.50) {};%
    \node (v6) at (2.00, -3.00) {};%
    \node (v7) at (0.00, -1.50) {};%
    \node (v8) at (4.00, -1.50) {};%
    
    \node [fit=(v1)(v7),line width=.5pt, fill=KITred!15,rounded corners=0.1cm] {};%
    \node [fit=(v2)(v3)(v4),line width=.5pt, fill=KITgreen!15,rounded corners=0.1cm] {};%
    \node [fit=(v5)(v6)(v8),line width=.5pt, fill=KITblue!15,rounded corners=0.1cm] {};%
    
    \path [edgeColor, line width=1pt]  (v1) edge node [above] {1} (v2);
    \path [edgeColor, line width=1pt]  (v1) edge node [left] {2} (v7);
    \path [edgeColor, line width=1pt]  (v2) edge node [above] {1} (v3);
    \path [edgeColor, line width=1pt]  (v2) edge node [left] {1} (v4);
    \path [edgeColor, line width=1pt]  (v2) edge node [left] {2} (v5);
    \path [edgeColor, line width=1pt]  (v3) edge node [right] {1} (v8);
    \path [edgeColor, line width=1pt]  (v5) edge node [right] {1} (v6);
    \path [edgeColor, line width=1pt]  (v5) edge node [above] {1} (v7);
    \path [edgeColor, line width=1pt]  (v5) edge node [above] {2} (v8);
    
    \node (v1v) at (v1) [vertex,draw=KITred,fill=KITred!15] {\gs};%
    \node (v2v) at (v2) [vertex,draw=KITgreen,fill=KITgreen!15] {\gs};%
    \node (v3v) at (v3) [vertex,draw=KITgreen,fill=KITgreen!15] {\gs};%
    \node (v4v) at (v4) [vertex,draw=KITgreen,fill=KITgreen!15] {\gs};%
    \node (v5v) at (v5) [vertex,draw=KITblue,fill=KITblue!15] {\gs};%
    \node (v6v) at (v6) [vertex,draw=KITblue,fill=KITblue!15] {\gs};%
    \node (v7v) at (v7) [vertex,draw=KITred,fill=KITred!15] {\gs};%
    \node (v8v) at (v8) [vertex,draw=KITblue,fill=KITblue!15] {\gs};%
\end{scope}
\end{tikzpicture}
\end{figure}

To represent the topology of the public transit network without its time dependency, we define the~\emph{layout graph}~$\layoutGraph$.
The set of \emph{connections} between a pair~$\stopA,\stopB$ of stops is given by
\begin{equation*}
\connections(\stopA,\stopB)\coloneqq\left\{ \tripSegment{\aTrip}{i}{i+1} \mid \aTrip\in\trips, \stopOfStopEvent{\aTrip}{i}=\stopA, \stopOfStopEvent{\aTrip}{i+1}=\stopB \right\} \cup \left\{ (\stopA,\stopB) \mid (\stopA,\stopB) \in \footpaths \right\}.
\end{equation*}
Thus, a connection is either a trip segment between two consecutive stops or a footpath.
Then the layout graph is defined as $\layoutGraph=\left(\stops, \layoutEdges, \layoutEdgeWeight\right)$, with the set of edges~$\layoutEdges \subseteq \stops \times \stops$ and edge weight function~$\layoutEdgeWeight:\layoutEdges \to \mathbb{N}$ defined by
\begin{align*}
\layoutEdges &\coloneqq \left\{(\stopA,\stopB) \mid \connections(\stopA,\stopB) \neq \emptyset \right\},\\
\layoutEdgeWeight\left((\stopA,\stopB)\right) &\coloneqq \absoluteVal{\connections(\stopA,\stopB)}.
\end{align*}
An illustration of a layout graph is given in Figure~\ref{fig:metisgraph}.
The stop partition~$\partition$ is obtained by generating the layout graph and running a graph partitioning algorithm of choice.
Due to the weight function, the partitioning algorithm will attempt to avoid separating stops which have many connections between them.

\subsection{Flag Computation}
\label{subsec:flag-computation}
To compute the flags, the full-range profile problem is solved for all pairs of source and target stops.
As with \algoname{TB-CST}, this is done by running one-to-all Profile-\tb search for every possible source stop.
After each \tb run of the Profile-\tb search, all newly found journeys are unpacked.
For a journey~$\aJourney$ to a target stop~$\targetStop$ and each transfer~$\aTransfer$ in~$\aJourney$, the flag~$\aFlag(\aTransfer,\partition(\targetStop))$ is set to~$1$.
Once all flags have been computed, transfers for which no flags are set to~$1$ can be removed from~$\transfers$.

\subsection{Optimizations}
\label{subsec:optimizations}

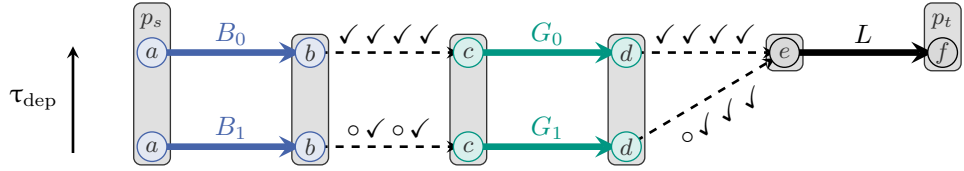
\begin{figure}[tp!]
    \caption{An example network illustrating the need for departure time fixing.
        Grey boxes represent stops.
        Nodes within the boxes represent stop events and are labeled with their indices along the respective trip.
        Within a stop, events are depicted in increasing order of time from bottom to top.
        Colored edges represent trips, with trips of the same line using the same color.
        Dashed edges with arrows represent transfers.
        Assume that the stop~$\targetStop$ is the only stop in cell~$i$ of the stop partition~$\partition$.
        Each transfer~$\aTransfer$ is labeled with checkmarks which indicate whether the flag of~$\aTransfer$ for cell~$i$ is set to 1 or not.
        From left to right, these represent various configurations of the flag computation algorithm: unmodified Profile-\tb, departure time buffering, flag augmentation, departure time buffering + flag augmentation.
    }
    \label{fig:departure-time-buffering}
    \centering
    \begin{tikzpicture}[scale=1.04]
\begin{scope}[yscale=0.8]
    \node (b0a) at (0.00, 0.00) {};%
    \node (b0b) at (2.00, 0.00) {};%
    \node (g0c) at (4.00, 0.00) {};%
    \node (g0d) at (6.00, 0.00) {};%
    \node (le) at  (8.00, 0.00) {};%
    \node (lf) at  (10.00, 0.00) {};%
    \node (b1a) at (0.00, -1.50) {};%
    \node (b1b) at (2.00, -1.50) {};%
    \node (g1c) at (4.00, -1.50) {};%
    \node (g1d) at (6.00, -1.50) {};%
    \node (source_label) at (0.00, 0.50) {};%
    \node (target_label) at (10.00, 0.50) {};%
    \node (arrow_begin) at (-1.00, -1.75) {};%
    \node (arrow_end) at (-1.00, 0.25) {};%
    
    \node [fit=(b0a)(b1a)(source_label),line width=.5pt, draw=nodeColor!100,fill=nodeColor!15,rounded corners=0.1cm] {};%
    \node [fit=(b0b)(b1b),line width=.5pt, draw=nodeColor!100,fill=nodeColor!15,rounded corners=0.1cm] {};%
    \node [fit=(g0c)(g1c),line width=.5pt, draw=nodeColor!100,fill=nodeColor!15,rounded corners=0.1cm] {};%
    \node [fit=(g0d)(g1d),line width=.5pt, draw=nodeColor!100,fill=nodeColor!15,rounded corners=0.1cm] {};%
    \node [fit=(le),line width=.5pt, draw=nodeColor!100,fill=nodeColor!15,rounded corners=0.1cm] {};%
    \node [fit=(lf)(target_label),line width=.5pt, draw=nodeColor!100,fill=nodeColor!15,rounded corners=0.1cm] {};%
    
    \draw [KITblue, line width=2.5pt, routeArrow] (b0a) -- (b0b);
    \draw [KITblue, line width=2.5pt, routeArrow] (b1a) -- (b1b);
    \draw [KITgreen, line width=2.5pt, routeArrow] (g0c) -- (g0d);
    \draw [KITgreen, line width=2.5pt, routeArrow] (g1c) -- (g1d);
    \draw [KITblack, line width=2.5pt, routeArrow] (le) -- (lf);
    
    \draw [black, line width=1pt, routeArrow] (arrow_begin) -- (arrow_end);
    \node [align=left] at ( -1.50, -0.75) {$\departureTime$};%

    \node [align=left,text=KITblue] at ( 1.00, 0.30) {$B_0$};%
    \node [align=left,text=KITblue] at ( 1.00, -1.20) {$B_1$};%
    \node [align=left,text=KITgreen] at ( 5.00, 0.30) {$G_0$};%
    \node [align=left,text=KITgreen] at ( 5.00, -1.20) {$G_1$};%
    \node [align=left,text=KITblack] at ( 9.00, 0.30) {$L$};%

    \draw [edgeColor, line width=1pt, dashed, edgeArrow]  (b0b) -- (g0c);
    \draw [edgeColor, line width=1pt, dashed, edgeArrow]  (b1b) -- (g1c);
    \draw [edgeColor, line width=1pt, dashed, edgeArrow]  (g0d) -- (le);
    \draw [edgeColor, line width=1pt, dashed, edgeArrow]  (g1d) -- (le);
    
    \node [align=left] at (3.00, 0.30) {\checkmark\,\checkmark\,\checkmark\,\checkmark};%
    \node [align=left] at (3.00, -1.20) {$\circ$\,\checkmark\,$\circ$\,\checkmark};%
    \node [align=left] at (7.00, 0.30) {\checkmark\,\checkmark\,\checkmark\,\checkmark};%
    \node [align=left] at (7.20, -1.00) {\rotatebox{30}{$\circ$\,\checkmark\,\checkmark\,\checkmark}};%
    
    \node (b0a_v) at (b0a) [vertex,draw=KITblue,fill=KITblue!15] {\gs};%
    \node (b1a_v) at (b1a) [vertex,draw=KITblue,fill=KITblue!15] {\gs};%
    \node (b0b_v) at (b0b) [vertex,draw=KITblue,fill=KITblue!15] {\gs};%
    \node (b1b_v) at (b1b) [vertex,draw=KITblue,fill=KITblue!15] {\gs};%
    \node (g0c_v) at (g0c) [vertex,draw=KITgreen,fill=KITgreen!15] {\gs};%
    \node (g1c_v) at (g1c) [vertex,draw=KITgreen,fill=KITgreen!15] {\gs};%
    \node (g0d_v) at (g0d) [vertex,draw=KITgreen,fill=KITgreen!15] {\gs};%
    \node (g1d_v) at (g1d) [vertex,draw=KITgreen,fill=KITgreen!15] {\gs};%
    \node (le_v)  at (le)  [vertex,draw=KITblack,fill=KITblack!15] {\gs};%
    \node (lf_v)  at (lf)  [vertex,draw=KITblack,fill=KITblack!15] {\gs};%

     \node at (b0a) [text=nodeColor!100] {\small{$a$}};%
     \node at (b1a) [text=nodeColor!100] {\small{$a$}};%
     \node at (b0b) [text=nodeColor!100] {\small{$b$}};%
     \node at (b1b) [text=nodeColor!100] {\small{$b$}};%
     \node at (g0c) [text=nodeColor!100] {\small{$c$}};%
     \node at (g1c) [text=nodeColor!100] {\small{$c$}};%
     \node at (g0d) [text=nodeColor!100] {\small{$d$}};%
     \node at (g1d) [text=nodeColor!100] {\small{$d$}};%
     \node at (le)  [text=nodeColor!100] {\small{$e$}};%
     \node at (lf)  [text=nodeColor!100] {\small{$f$}};%
     \node at (source_label) [text=nodeColor!100] {\small{$\sourceStop$}};%
     \node at (target_label) [text=nodeColor!100] {\small{$\targetStop$}};%
\end{scope}
\end{tikzpicture}
\end{figure}

\subparagraph*{Departure Time Buffering.}
Due to line pruning, a \tb query always enters the earliest reachable trip of a line; later trips of the same line are not explored.
However, because Profile-\tb processes departure times in decreasing order and applies self-pruning, it returns journeys which depart as late as possible.
These two pruning rules conflict, leading to situations where \arcflagtb fails to find a Pareto-optimal journey.
An example of this is shown in Figure~\ref{fig:departure-time-buffering}.
An unmodified Profile-\tb search from~$\sourceStop$ will find the journey~$\aJourney_0\coloneqq\langle\tripSegment{B_0}{a}{b},\tripSegment{G_0}{c}{d},\tripSegment{L}{e}{f}\rangle$ and flag it for cell~$i$.
However, it will not flag the journey~$\aJourney_1\coloneqq\langle\tripSegment{B_1}{a}{b},\tripSegment{G_1}{c}{d},\tripSegment{L}{e}{f}\rangle$, which has an earlier departure time and is therefore processed in a later run, but has the same arrival time and number of trips.
An \arcflagtb query from~$\sourceStop$ to~$\targetStop$ with departure time~$\deptime{B_1}{a}$ will enter~$B_1$ but not relax the unflagged transfer~$\transfer{B_1}{b}{G_1}{c}$.
While~$\transfer{B_0}{b}{G_0}{c}$ is flagged, the query will not enter~$B_0$ due to line pruning and therefore not relax this transfer either.

To solve this issue, we introduce the notion of the itinerary.
An~\emph{itinerary} is a generalized description of a journey which specifies the lines used and the stop indices where they are entered and exited, but not the trips used.
Corresponding to a stop event~$\stopEvent{\aTrip}{i}$ is the \emph{line event}~$\stopEvent{\aLine}{i}$ where~$\aTrip\in\trips(\aLine)$.
The stop visited by~$\stopEvent{\aLine}{i}$ is denoted as~$\stopOfStopEvent{\aLine}{i}$.
A trip segment~$\tripSegment{\aTrip}{i}{j}$ corresponds to the \emph{line segment}~$\tripSegment{\aLine}{i}{j}$.
An itinerary is therefore a sequence of line segments.
The itinerary describing a journey~$\aJourney=\langle\tripSegment{\aTrip_1}{b_1}{e_1},\dots,\tripSegment{\aTrip_k}{b_k}{e_k}\rangle$ is given by~$\anItinerary(\aJourney)=\langle\lineSegment{\aLine_1}{b_1}{e_1},\dots,\lineSegment{\aLine_k}{b_k}{e_k}\rangle$, where~$\aTrip_i\in\trips(\aLine_i)$ for~$1 \leq i \leq k$.
For a line~$\aLine$, an index~$i$ and a departure time~$\departureTime$, let~$\bufferedTrip(\aLine,i,\departureTime)$ denote the earliest trip of~$\aLine$ which departs at~$\stopOfStopEvent{\aLine}{i}$ no earlier than~$\departureTime$.
For an itinerary~$\anItinerary=\langle\lineSegment{\aLine_1}{b_1}{e_1},\dots,\lineSegment{\aLine_k}{b_k}{e_k}\rangle$, the journey~$\bufferedJourney(\anItinerary,\departureTime)$ is the journey with itinerary~$\anItinerary$ which takes the earliest reachable trip of every line when starting with departure time~$\departureTime$.
Formally, $\bufferedJourney(\anItinerary,\departureTime)=\langle\tripSegment{\aTrip_1}{b_1}{e_1},\dots,\tripSegment{\aTrip_k}{b_k}{e_k}\rangle$ with~$\aTrip_i=\bufferedTrip(\aLine_i,b_i,\departureTime^i)$ and
\[\departureTime^i=\begin{cases}
\departureTime+\transfertime{\sourceStop}{\stopOfStopEvent{\aLine_1}{b_1}} & \text{if } i=1,\\
\arrtime{\aTrip_{i-1}}{e_{i-1}}+\transfertime{\stopOfStopEvent{\aLine_{i-1}}{e_{i-1}}}{\stopOfStopEvent{\aLine_i}{b_i}} & \text{otherwise.}
\end{cases}
\]

In Figure~\ref{fig:departure-time-buffering}, $\aJourney_0$ and~$\aJourney_1$ have the same itinerary~$\anItinerary$.
To ensure that the query from~$\sourceStop$ to~$\targetStop$ with departure time~$\deptime{B_1}{a}$ is answered correctly by \arcflagtb, $\bufferedJourney(\anItinerary,\deptime{B_1}{a})=\aJourney_1$ must be flagged as well.
In general, consider the one-to-all Profile-\tb search from a source stop~$\sourceStop$.
For a \tb run with departure time~$\departureTime$ and a target stop~$\targetStop$, let~$\paretoset$ be the found Pareto set.
We define the \emph{buffered Pareto set}
\[\bufferedParetoSet(\paretoset,\departureTime)\coloneqq\{ \bufferedJourney(\anItinerary(\aJourney),\departureTime) \mid \aJourney\in\paretoset \}.\]
Since~$\paretoset$ is a Pareto set, we know that every journey~$\aJourney\in\paretoset$ has the same arrival time as~$\bufferedJourney(\aJourney,\departureTime)$.
Hence, the last trip segment of~$\bufferedJourney(\anItinerary(\aJourney),\departureTime)$ and~$\aJourney$ is always identical.
However, for the other trip segments, $\bufferedJourney(\anItinerary(\aJourney),\departureTime)$ may use earlier trips than~$\aJourney$.
We modify the Profile-\tb search so that it flags all transfers in~$\bufferedParetoSet(\paretoset,\departureTime)$.
To do this efficiently, we employ an approach which we call~\emph{departure time buffering}.
An itinerary~$\anItinerary$ beginning with the line segment~$\lineSegment{\aLine_1}{b_1}{e_1}$ is~\emph{unpacked} within the interval~$(\uptau_1,\uptau_2]$ as follows:
For a trip~$\aTrip_1\in\trips(\aLine_1)$, let~$\departureTime(\anItinerary,\aTrip_1)\coloneqq\deptime{\aTrip_1}{b_1}-\transfertime{\sourceStop}{\stopOfStopEvent{\aTrip_1}{b_1}}$ be the departure time of a journey with the itinerary~$\anItinerary$ that uses~$\aTrip_1$ as the first trip.
For each trip~$\aTrip_1\in\trips(\aLine_1)$ with~$\departureTime(\anItinerary,\aTrip_1)\in(\uptau_1,\uptau_2]$, the journey~$\bufferedJourney(\anItinerary, \departureTime(\anItinerary,\aTrip_1))$ is constructed and its transfers are flagged.

For each stop~$\aStop$ and round~$n$, the algorithm maintains not only the earliest arrival time~$\arrivalTime(\aStop,n)$ but a buffered itinerary~$\anItinerary(\aStop,n)$, which represents the journey associated with~$\arrivalTime(\aStop,n)$, as well as the departure time~$\departureTime(\aStop,n)$ of the run in which~$\arrivalTime(\aStop,n)$ was last changed.
If~$\arrivalTime(\aStop,n)$ is improved during a run with departure time~$\departureTime$, then the journey corresponding to this arrival time is not flagged right away.
Instead, after the end of the run, the algorithm unpacks the buffered itinerary~$\anItinerary(\aStop,n)$ within the interval~$(\departureTime,\departureTime(\aStop,n))$ (unless~$\anItinerary(\aStop,n)$ has not been set before).
Afterwards, the buffered itinerary~$\anItinerary(\aStop,n)$ is updated by unpacking the journey corresponding to the new value of~$\arrivalTime(\aStop,n)$.
After the last \tb run of the profile search, the remaining buffered itineraries are processed.
For every stop~$\aStop$ and round~$n$ such that~$\arrivalTime(\aStop,n)<\infty$, the itinerary~$\anItinerary(\aStop,n)$ is unpacked within the interval~$(-\infty,\departureTime(\aStop,n))$.

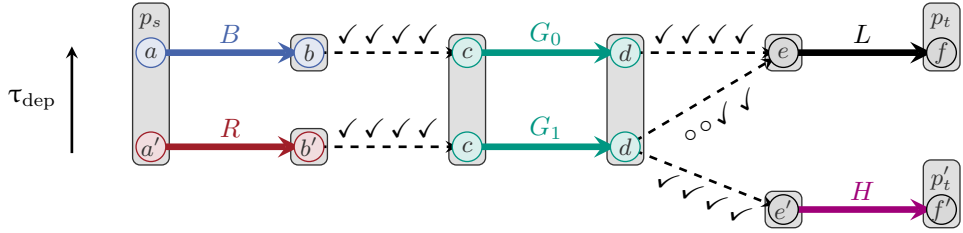
\begin{figure}[tp!]
    \caption{An example network illustrating the need for flag augmentation.
    Nodes, edges and checkmarks have the same meaning as in Figure~\ref{fig:departure-time-buffering}.
    Assume that the stops~$\targetStop$ and~$\targetStop'$ are the only stops in cell~$i$ of the stop partition~$\partition$.}
    \label{fig:flag-augmentation}
    \centering
    \begin{tikzpicture}[scale=1.04]
\begin{scope}[yscale=0.8]
    \node (b0a) at (0.00, 0.00) {};%
    \node (b0b) at (2.00, 0.00) {};%
    \node (g0c) at (4.00, 0.00) {};%
    \node (g0d) at (6.00, 0.00) {};%
    \node (le) at  (8.00, 0.00) {};%
    \node (lf) at  (10.00, 0.00) {};%
    \node (he) at  (8.00, -2.50) {};%
    \node (hf) at  (10.00, -2.50) {};%
    \node (b1a) at (0.00, -1.50) {};%
    \node (b1b) at (2.00, -1.50) {};%
    \node (g1c) at (4.00, -1.50) {};%
    \node (g1d) at (6.00, -1.50) {};%
    \node (source_label) at (0.00, 0.50) {};%
    \node (target_label) at (10.00, 0.50) {};%
    \node (target2_label) at (10.00, -2.00) {};%
    \node (arrow_begin) at (-1.00, -1.75) {};%
    \node (arrow_end) at (-1.00, 0.25) {};%
    
    \node [fit=(b0a)(b1a)(source_label),line width=.5pt, draw=nodeColor!100,fill=nodeColor!15,rounded corners=0.1cm] {};%
    \node [fit=(b0b),line width=.5pt, draw=nodeColor!100,fill=nodeColor!15,rounded corners=0.1cm] {};%
    \node [fit=(b1b),line width=.5pt, draw=nodeColor!100,fill=nodeColor!15,rounded corners=0.1cm] {};%
    \node [fit=(g0c)(g1c),line width=.5pt, draw=nodeColor!100,fill=nodeColor!15,rounded corners=0.1cm] {};%
    \node [fit=(g0d)(g1d),line width=.5pt, draw=nodeColor!100,fill=nodeColor!15,rounded corners=0.1cm] {};%
    \node [fit=(le),line width=.5pt, draw=nodeColor!100,fill=nodeColor!15,rounded corners=0.1cm] {};%
    \node [fit=(lf)(target_label),line width=.5pt, draw=nodeColor!100,fill=nodeColor!15,rounded corners=0.1cm] {};%
    \node [fit=(he),line width=.5pt, draw=nodeColor!100,fill=nodeColor!15,rounded corners=0.1cm] {};%
    \node [fit=(hf)(target2_label),line width=.5pt, draw=nodeColor!100,fill=nodeColor!15,rounded corners=0.1cm] {};%
    
    \draw [KITblue, line width=2.5pt, routeArrow] (b0a) -- (b0b);
    \draw [KITred, line width=2.5pt, routeArrow] (b1a) -- (b1b);
    \draw [KITgreen, line width=2.5pt, routeArrow] (g0c) -- (g0d);
    \draw [KITgreen, line width=2.5pt, routeArrow] (g1c) -- (g1d);
    \draw [KITblack, line width=2.5pt, routeArrow] (le) -- (lf);
    \draw [KITlilac, line width=2.5pt, routeArrow] (he) -- (hf);
    
    \draw [black, line width=1pt, routeArrow] (arrow_begin) -- (arrow_end);
    \node [align=left] at ( -1.50, -0.75) {$\departureTime$};%

    \node [align=left,text=KITblue] at ( 1.00, 0.30) {$B$};%
    \node [align=left,text=KITred] at ( 1.00, -1.20) {$R$};%
    \node [align=left,text=KITgreen] at ( 5.00, 0.30) {$G_0$};%
    \node [align=left,text=KITgreen] at ( 5.00, -1.20) {$G_1$};%
    \node [align=left,text=KITblack] at ( 9.00, 0.30) {$L$};%
    \node [align=left,text=KITlilac] at ( 9.00, -2.20) {$H$};%

    \draw [edgeColor, line width=1pt, dashed, edgeArrow]  (b0b) -- (g0c);
    \draw [edgeColor, line width=1pt, dashed, edgeArrow]  (b1b) -- (g1c);
    \draw [edgeColor, line width=1pt, dashed, edgeArrow]  (g0d) -- (le);
    \draw [edgeColor, line width=1pt, dashed, edgeArrow]  (g1d) -- (le);
    \draw [edgeColor, line width=1pt, dashed, edgeArrow]  (g1d) -- (he);
    
    \node [align=left] at (3.00, 0.30) {\checkmark\,\checkmark\,\checkmark\,\checkmark};%
    \node [align=left] at (3.00, -1.20) {\checkmark\,\checkmark\,\checkmark\,\checkmark};%
    \node [align=left] at (7.00, 0.30) {\checkmark\,\checkmark\,\checkmark\,\checkmark};%
    \node [align=left] at (7.20, -1.00) {\rotatebox{30}{$\circ$\,$\circ$\,\checkmark\,\checkmark}};%
    \node [align=left] at (7.00, -2.30) {\rotatebox{-25}{\checkmark\,\checkmark\,\checkmark\,\checkmark}};%
    
    \node (b0a_v) at (b0a) [vertex,draw=KITblue,fill=KITblue!15] {\gs};%
    \node (b1a_v) at (b1a) [vertex,draw=KITred,fill=KITred!15] {\gs};%
    \node (b0b_v) at (b0b) [vertex,draw=KITblue,fill=KITblue!15] {\gs};%
    \node (b1b_v) at (b1b) [vertex,draw=KITred,fill=KITred!15] {\gs};%
    \node (g0c_v) at (g0c) [vertex,draw=KITgreen,fill=KITgreen!15] {\gs};%
    \node (g1c_v) at (g1c) [vertex,draw=KITgreen,fill=KITgreen!15] {\gs};%
    \node (g0d_v) at (g0d) [vertex,draw=KITgreen,fill=KITgreen!15] {\gs};%
    \node (g1d_v) at (g1d) [vertex,draw=KITgreen,fill=KITgreen!15] {\gs};%
    \node (le_v)  at (le)  [vertex,draw=KITblack,fill=KITblack!15] {\gs};%
    \node (lf_v)  at (lf)  [vertex,draw=KITblack,fill=KITblack!15] {\gs};%
    \node (he_v)  at (he)  [vertex,draw=KITblack,fill=KITblack!15] {\gs};%
    \node (hf_v)  at (hf)  [vertex,draw=KITblack,fill=KITblack!15] {\gs};%

     \node at (b0a) [text=nodeColor!100] {\small{$a$}};%
     \node at (b1a) [text=nodeColor!100] {\small{$a'$}};%
     \node at (b0b) [text=nodeColor!100] {\small{$b$}};%
     \node at (b1b) [text=nodeColor!100] {\small{$b'$}};%
     \node at (g0c) [text=nodeColor!100] {\small{$c$}};%
     \node at (g1c) [text=nodeColor!100] {\small{$c$}};%
     \node at (g0d) [text=nodeColor!100] {\small{$d$}};%
     \node at (g1d) [text=nodeColor!100] {\small{$d$}};%
     \node at (le)  [text=nodeColor!100] {\small{$e$}};%
     \node at (lf)  [text=nodeColor!100] {\small{$f$}};%
     \node at (he)  [text=nodeColor!100] {\small{$e'$}};%
     \node at (hf)  [text=nodeColor!100] {\small{$f'$}};%
     \node at (source_label) [text=nodeColor!100] {\small{$\sourceStop$}};%
     \node at (target_label) [text=nodeColor!100] {\small{$\targetStop$}};%
     \node at (target2_label) [text=nodeColor!100] {\small{$\targetStop'$}};%
\end{scope}
\end{tikzpicture}
\end{figure}

\subparagraph*{Flag Augmentation.}
Departure time buffering does not fix all issues caused by the incompatibility of line pruning and self-pruning.
Consider the example shown in Figure~\ref{fig:flag-augmentation}.
Once again, an unmodified Profile-\tb search from~$\sourceStop$ will find the journey~$\aJourney_0\coloneqq\langle\tripSegment{B}{a}{b},\tripSegment{G_0}{c}{d},\tripSegment{L}{e}{f}\rangle$ and flag it for cell~$i$, whereas the equivalent journey~$\aJourney_1\coloneqq\langle\tripSegment{R}{a'}{b'},\tripSegment{G_1}{c}{d},\tripSegment{L}{e}{f}\rangle$ is discarded.
In this case, however, departure time buffering will not cause~$\aJourney_1$ to be flagged either because it starts with a different line than~$\aJourney_0$.
Once again, consider an \arcflagtb query from~$\sourceStop$ to~$\targetStop$ with departure time~$\deptime{R}{a'}$.
If the transfer~$\transfer{R}{b'}{G_1}{c}$ is not flagged, then the algorithm will enter~$G_0$ and find~$\aJourney_0$.
However, in the example network this transfer is flagged due to another journey~$\aJourney_1'\coloneqq\langle\tripSegment{R}{a'}{b'},\tripSegment{G_1}{c}{d},\tripSegment{H}{e'}{f'}\rangle$, which leads another target stop~$\targetStop'$ in cell~$i$.
As a consequence, $G_1$ is entered but the unflagged transfer~$\transfer{G_1}{d}{L}{e}$ is not relaxed, while~$G_0$ is not entered due to line pruning.

To fix this issue, we define the \emph{augmented flags function}~$\aFixedFlag : \transfers \times \left\{1, \dots, \numCells\right\} \to \left\{0, 1\right\}$.
Consider a line~$\aLine$, a trip~$\tripA \in \trips(\aLine)$ and a transfer~$\aTransfer=\transfer{\tripA}{i}{\tripB}{j} \in \transfers$.
We define the set of \emph{successor transfers} $\successorTransfers(\aTransfer)$ as
\begin{align*}
\successorTransfers(\transfer{\tripA}{i}{\tripB}{j})&\coloneqq\{ \transfer{\tripA'}{i}{\tripB}{j} \in \transfers \mid \tripA' \in \trips\left(\aLine\right), \tripA \preceq \tripA' \}.
\end{align*}
Then~$\aFixedFlag$ is defined as follows for a transfer~$\aTransfer\in\transfers$ and cell~$i$:

\begin{equation*}
	\aFixedFlag(\aTransfer, i) \coloneqq \bigvee\limits_{\aTransfer' \in \successorTransfers(\aTransfer)} \aFlag(\aTransfer, i)
\end{equation*}

In the example from Figure~\ref{fig:flag-augmentation}, using~$\aFixedFlag$ instead of~$\aFlag$ resolves the problem, provided that all transfers which occur in~$\aJourney_1$ are included in the set~$\transfers$ of transfers generated by the \tb preprocessing phase.
Note that flag augmentation alone (without departure time buffering) will not fix the issue shown in Figure~\ref{fig:departure-time-buffering} because the transfer~$\transfer{B_1}{b}{G_1}{c}$ will not be flagged.
Thus, both fixes must be combined.

Unfortunately, the rules for pruning unnecessary transfers which are employed by the \tb preprocessing phase are too strong to guarantee that the required transfers are always generated.
It is possible to construct examples akin to Figure~\ref{fig:departure-time-buffering} where the transfer~$\transfer{G_1}{d}{L}{f}$ is not included in~$\transfers$, for example because a transfer to a different trip than~$L$ is preferred.
In these cases, both \arcflagtb and \algoname{TB-CST} will fail to return correct results.
Adapting the pruning rules of the \tb preprocessing phase to resolve these issues remains an open problem.

\subparagraph*{Flag Compression.}
For an edge~$\anEdge$, we call the set of flags~$\aFlag(\anEdge,i)$ for~$1 \leq i \leq k$ its~\emph{flag pattern}.
Bauer \etal~\cite{Bau09} observed for \algoname{Arc-Flags} on road networks that many edges in the graph share the same flag pattern.
They exploit this with the following compression technique:
All flag patterns which occur in the graph are stored in a global array~$A$.
For each edge~$\anEdge$, the algorithm does not store the flag pattern of~$\anEdge$ directly, but rather the index~$i$ for which~$A[i]$ holds the flag pattern of~$\anEdge$.
This significantly reduces memory consumption at the cost of an additional pointer access whenever an edge is relaxed.
We also apply this compression technique in \arcflagtb and sort the flag pattern array in decreasing order of occurrence.
This ensures that the most commonly accessed flag patterns are stored close together in memory, which increases the likelihood of cache hits.

\subsection{Comparison}
\label{subsec:arcflagtb-comparison}
We conclude this section by comparing \arcflagtb to similar approaches.

\subparagraph*{TB-CST.}
\algoname{TB-CST} (without split trees) stores a generalized shortest path tree for every possible source stop.
This offers a near-perfect reduction in the query search space but at the expense of quadratic memory consumption.
The memory consumption of \arcflagtb is in~$\Theta(\absoluteVal{\transfers}\numCells)$, where~$\numCells$ is the number of cells.
Thus, \arcflagtb can be seen as a way to interpolate between \tb and \algoname{TB-CST} in terms of query search space and memory consumption.
For~$\numCells=1$, every non-superfluous transfer will be flagged, and thus the search space will be identical to that of \tb with a minimal set of transfers.
For~$\numCells=\numVertices$, the flags provide perfect information about whether a transfer is required to reach the target node.

An advantage of our approach is that the transfer flags provide information about which specific trips should be entered, whereas the \algoname{TB-CST} search graph only provides information about entire lines.
This means that \arcflagtb does not have to invest additional effort during the query phase in order to find the earliest reachable trip of each line.

\subparagraph*{Time-Expanded Arc-Flags.}
Conceptually, our approach is similar to \algoname{Arc-Flags} on a time-expanded graph, albeit with \tb as a query algorithm instead of \algoname{Dijkstra's Algorithm}.
Delling \etal~\cite{Del09c} observed low speedups when applying \algoname{Arc-Flags} to time-expanded graphs.
We analyze some of the issues causing this and how \arcflagtb overcomes them.
In a time-expanded graph, each visit of a vehicle at a stop is modeled with three nodes: an \emph{arrival node}, a \emph{transfer node} and a \emph{departure node}.
A journey corresponds to a path between two transfer nodes.
However, boundary nodes in the partition may also be departure or arrival nodes.
Consider for example a boundary node~$\aVertex$ of cell~$i$ which is an arrival node corresponding to the stop event~$\stopEvent{\aTrip}{i}$.
\algoname{Arc-Flags} will compute and flag a backward shortest-path tree rooted in~$\aVertex$.
A path in this tree corresponds to a ``journey'' which ends with the passenger remaining seated in~$\aTrip$.
However, there is no guarantee that this path can be extended to a Pareto-optimal journey which ends at a transfer node in cell~$i$.
It is possible that entering~$\aTrip$ is never required to enter cell~$i$.
In this case, \algoname{Arc-Flags} produces superfluous flags.
\arcflagtb avoids this problem by performing a one-to-all profile search from all stops, including those which are not boundary nodes in the layout graph.
While this requires~$\Omega(\absoluteVal{\stops}^2)$ preprocessing time, it reduces the number of set flags considerably.

Another feature of \arcflagtb which reduces the search space is that it flags transfers between stop events.
In the time-expanded graph, a transfer corresponds to an entire path between an arrival and a departure node, which may pass through several transfer nodes.
Consider two flagged transfers~$\transfer{T_1}{i_1}{T_2}{i_2}$ and~$\transfer{T_3}{i_3}{T_4}{i_4}$ whose paths in the time-expanded graph intersect.
This has the effect of creating ``virtual'' transfers~$\transfer{T_1}{i_1}{T_4}{i_4}$ and~$\transfer{T_3}{i_3}{T_2}{i_2}$, which may not be flagged.
\algoname{Arc-Flags} on the time-expanded graph will explore these transfers, whereas \arcflagtb will not.
Note that \arcflagtb only flags transfers, not trip segments.
This is because flagging trip segments would not provide any benefit beyond the first round of an \arcflagtb query:
If a trip segment is not flagged for a specific cell, then neither are its incoming or outgoing transfers.
Thus, an unflagged trip segment can only be entered during the first round, and no further trip segments are reachable from there.

Finally, Delling \etal~note that all paths in a time-expanded graph have optimal arrival time, and that a speedup is only achieved with suitable tiebreaking choices between equivalent paths.
They observed that their tiebreaking choices conflicted with their implementation of line pruning, Node-Blocking.
In \arcflagtb, the tiebreaking choices are dictated by the self-pruning of Profile-\tb.
We also observed conflicts with line pruning, but we analyzed these issues in detail and resolved them.
This allows \arcflagtb to fully benefit from both pruning rules, unlike previous approaches.

\section{Experimental Evaluation}
\label{sec:experiments}
We evaluate the performance of \arcflagtb on a selection of real-world public transit networks.
All experiments were run on a machine equipped with an \codestyling{AMD EPYC 7702P} CPU with 64 cores and 1\,TB of RAM. 
Code for \tb and \arcflagtb was written in \codestyling{C++} and compiled using GCC with optimizations enabled (\codestyling{-march=native -O3}).
For \algoname{TB-CST}, we used the original code provided to us by the author~\cite{Wit15}.
The preprocessing phases of \algoname{TB-CST} and \arcflagtb, which run one-to-all Profile-\tb from each stop, were parallelized with 64 threads.

Our datasets are taken from GTFS feeds of the public transit networks of Germany\footnote{\url{https://gtfs.de/}}, Paris\footnote{\textcopyright\;\url{https://navitia.io/}}, Sweden\footnote{\url{https://trafiklab.se/}} and Switzerland\footnote{\url{https://opentransportdata.swiss/}}.
Details are listed in Table~\ref{tab:datasets}.
For each network, we extracted the timetable of two consecutive weekdays in order to allow for overnight journeys.
\begin{table}[tb!]
	\caption{An overview of the networks on which we performed our experiments.
        Stops, lines, trips and footpaths are from the GTFS datasets.
        Transfers were generated by the \tb precomputation.}
	\label{tab:datasets}
	\vspace*{5mm}
	\centering
	\begin{tabular}{lrrrrr}
		\toprule
		Network & Stops & Lines & Trips & Footpaths & Transfers \\
		\midrule
		Germany & \numprint{441465} & \numprint{207801} & \numprint{1559118} & \numprint{1172464} & \numprint{60919877} \\
		Paris & \numprint{41757} & \numprint{9558} & \numprint{215526} & \numprint{445912} & \numprint{23284123} \\ %
		Sweden & \numprint{48007} & \numprint{15627} & \numprint{248977} & \numprint{2118} & \numprint{14771466} \\ %
		Switzerland & \numprint{30861} & \numprint{18235} & \numprint{559752} & \numprint{20864} & \numprint{9142826} \\ %
		\bottomrule
	\end{tabular}
\end{table}

\begin{table}[t!]
	\caption{Performance of \arcflagtb depending on the number of cells~$\numCells$.
		Departure time fixing and flag augmentation are enabled for all experiments.
		Query times and success rates are averaged over~\numprint{10000} random queries.
		Success rate for queries is the percentage of queries for which \arcflagtb found a correct Pareto set.
		For journeys, it is the percentage of journeys in the correct Pareto sets for which \arcflagtb found an equivalent journey.
		Query times and memory consumption are measured with and without flag compression.
		Note that the preprocessing time does not include the partitioning, which was limited to 10 minutes in all configurations.
		Due to time constraints, we only report the configuration~$\numCells=\numprint{1024}$ for Germany.
		Query times for~$\numCells=1$~are~for~\tb.}
	\label{tab:results}
	\vspace*{5mm}
	\centering
	\begin{tabular*}{\textwidth}{@{\,}l@{\extracolsep{\fill}}r@{\extracolsep{\fill}}r@{\extracolsep{\fill}}r@{\extracolsep{\fill}}r@{\extracolsep{\fill}}r@{\extracolsep{\fill}}r@{\extracolsep{\fill}}r@{\extracolsep{\fill}}r@{\extracolsep{\fill}}r@{\extracolsep{\fill}}r@{\,}}
		\toprule
		\multirow{2}{*}{Network} & \multirow{2}{*}{$\numCells$} & \multirow{2}{*}{\thead{Prepro.\\ $\left[ \mathrm{hh}{:}\mathrm{mm}{:}\mathrm{ss}\right]$}} & \multicolumn{2}{c}{Query time $[\mu s]$} & \multicolumn{2}{c}{Memory $[\mathrm{MB}]$} & \multicolumn{2}{c}{Success rate $[\%]$} &  \\
		\cmidrule(){4-5} \cmidrule(){6-7} \cmidrule(){8-9}
		& & & Uncomp. & Comp. & Uncomp. &  Comp. & Queries & Journeys \\
		\midrule
		\multirow{2}{*}{Germany} & \numprint{1} & -- & \numprint{105809} & -- & -- & -- & -- & -- \\
		& \numprint{1024} & \printTime{37}{24}{17} & \numprint{739} & \numprint{1068} & \numprint{5120} & \numprint{725} & 93.53 & 95.12 \\[5pt]
		\multirow{6}{*}{Paris} & \numprint{1}	& -- & \numprint{4502} & --	& --	& -- & -- & -- \\
		& \numprint{64}	& \printTime{0}{37}{15}	& \numprint{865}	& \numprint{1528} 	& \numprint{450}	 	& \numprint{95} & 99.17 & 99.30 \\ 
		& \numprint{128}	& \printTime{0}{37}{19}	& \numprint{671}	& \numprint{1486}	& \numprint{513}	 	& \numprint{133} & 99.08 & 99.25	\\
		& \numprint{256}	& \printTime{0}{37}{29}	& \numprint{502}	& \numprint{1230}	& \numprint{639}	 	& \numprint{187}	& 98.98 & 99.18 \\
		& \numprint{512}	&\printTime{0}{37}{35}	& \numprint{393}	& \numprint{982} & \numprint{891}	 	& \numprint{282}	& 98.76 & 99.03 \\
		& \numprint{1024}	& \printTime{0}{37}{45}	& \numprint{331}	& \numprint{757} & \numprint{1434}	 	& \numprint{468}	& 98.43 & 98.78 \\[5pt]
		\multirow{6}{*}{Sweden} & \numprint{1} & -- & \numprint{7583} & -- & -- & -- & -- & -- \\
		& \numprint{64} & \printTime{0}{17}{24} & \numprint{265} & \numprint{288} & \numprint{376} & \numprint{60} & 95.77 & 96.61 \\
		& \numprint{128} & \printTime{0}{17}{26} & \numprint{167} & \numprint{202} & \numprint{428} & \numprint{69} & 95.36 & 96.24 \\
		& \numprint{256} & \printTime{0}{17}{27} & \numprint{121} & \numprint{164} & \numprint{534} & \numprint{84} & 95.14 & 96.01 \\
		& \numprint{512} & \printTime{0}{17}{30} & \numprint{97} & \numprint{140} & \numprint{744} & \numprint{116} & 94.86 & 95.82 \\
		& \numprint{1024} & \printTime{0}{17}{32} & \numprint{88} & \numprint{127} & \numprint{1229} & \numprint{184} & 94.37 & 95.49 \\[5pt]
		\multirow{6}{*}{Switzerland} & \numprint{1} & -- & \numprint{7043} & -- & -- & -- & -- & -- \\
		& \numprint{64} & \printTime{0}{13}{07} & \numprint{223} & \numprint{225} & \numprint{222} & \numprint{35} & 96.74 & 97.53 \\
		& \numprint{128} & \printTime{0}{13}{08} & \numprint{154} & \numprint{172} & \numprint{253} & \numprint{40} & 96.29 & 97.23 \\
		& \numprint{256} & \printTime{0}{13}{09} & \numprint{112} & \numprint{136} & \numprint{315} & \numprint{50} & 95.75 & 96.86 \\
		& \numprint{512} & \printTime{0}{13}{13} & \numprint{91} & \numprint{118} & \numprint{440} & \numprint{70} & 95.29 & 96.60 \\
		& \numprint{1024} & \printTime{0}{13}{14} & \numprint{81} & \numprint{108} & \numprint{698} & \numprint{114} & 94.80 & 96.29 \\
		\bottomrule
	\end{tabular*}
\end{table}

\subparagraph*{Partitioning.}
\label{subsec:partitioning} 
We use the \algoname{KaHIP}\footnote{\url{https://github.com/KaHIP/KaHIP}} \cite{San13} open-source graph partitioning library to partition our networks. 
\algoname{KaHIP} is based on a multilevel approach, \ie the input graph is coarsened, initially partitioned, and locally improved during uncoarsening.
In our experiments, overall better results are obtained when coarsening is computed using clustering rather than edge matching as usual.
More specifically, we use the memetic algorithm \codestyling{kaffpaE} with the strong social configuration and an imbalance parameter of $5\%$ in all of our experiments.
As a time limit, we set 10 minutes for all networks regardless of the number~$\numCells$ of desired cells.
In our experiments, higher time limits did not significantly improve the results regarding the total number of flags set and average query times. 

\begin{figure}[t!]
\caption{Average speedup over \tb and memory consumption of \arcflagtb (with and without flag compression) on the Paris and Switzerland networks, depending on the number~of~cells~$\numCells$.}
\label{fig:resultplot}
\centering
\resizebox{\textwidth}{!}{
\begin{tikzpicture}
    \begin{groupplot}[group style={group size= 2 by 2, vertical sep=60, horizontal sep=60}]            
        \nextgroupplot[title=\textbf{Paris -- Speedup}, xlabel={$\numCells$}, ylabel={Speedup}, symbolic x coords={64, 128, 256, 512, 1024}, ymin=0, ymax=15, legend cell align={left}, bar width = 8pt]
        \addplot[ybar, /pgf/bar shift=-6pt, fill=darkgreen, darkgreen, thick] table [x index=0, y index=1, col sep=comma] {csv/paris.csv};
        \addplot[ybar, /pgf/bar shift=6pt, fill=blue, blue, thick] table [x index=0, y index=2, col sep=comma] {csv/paris.csv};
            
        \nextgroupplot[title=\textbf{Paris -- Memory}, xlabel={$\numCells$}, ylabel={Memory $[\mathrm{MB}]$}, symbolic x coords={64, 128, 256, 512, 1024}, ymin=0, ymax=1500, legend cell align={left}, bar width = 8pt]
        \addplot[ybar, /pgf/bar shift=-6pt, fill=darkgreen, darkgreen, thick] table [x index=0, y index=3, col sep=comma] {csv/paris.csv};
        \addplot[ybar, /pgf/bar shift=6pt, fill=blue, blue, thick] table [x index=0, y index=4, col sep=comma] {csv/paris.csv};
        
        \nextgroupplot[title=\textbf{Switzerland -- Speedup}, xlabel={$\numCells$}, ylabel={Speedup}, symbolic x coords={64, 128, 256, 512, 1024}, ymin=0, ymax=100, legend cell align={left}, bar width = 8pt]
        \addplot[ybar, /pgf/bar shift=-6pt, fill=darkgreen, darkgreen, thick] table [x index=0, y index=1, col sep=comma] {csv/switzerland.csv};
        \addplot[ybar, /pgf/bar shift=6pt, fill=blue, blue, thick] table [x index=0, y index=2, col sep=comma] {csv/switzerland.csv};
        \coordinate (c1) at (rel axis cs:0,1);
        
        \nextgroupplot[title=\textbf{Switzerland -- Memory}, xlabel={$\numCells$}, ylabel={Memory $[\mathrm{MB}]$}, legend to name=sweden, symbolic x coords={64, 128, 256, 512, 1024}, ymin=0, ymax=800, legend cell align={left}, bar width = 8pt, legend image code/.code={
            \draw [#1] (0cm,-0.1cm) rectangle (0.2cm,0.25cm); }]
        \addplot[ybar, /pgf/bar shift=-6pt, fill=darkgreen, darkgreen, thick] table [x index=0, y index=3, col sep=comma] {csv/switzerland.csv};
        \addlegendentry{\arcflagtb (no flag compression)};
        \addplot[ybar, /pgf/bar shift=6pt, fill=blue, blue, thick] table [x index=0, y index=4, col sep=comma] {csv/switzerland.csv};
        \addlegendentry{\arcflagtb (flag compression)};
        \coordinate (c2) at (rel axis cs:1,1);
    \end{groupplot}
    \coordinate (c3) at ($(c1)!.5!(c2)$);
    \node[below] at (c3 |- current bounding box.south)
      {\pgfplotslegendfromname{sweden}};
\end{tikzpicture}
}
\end{figure}
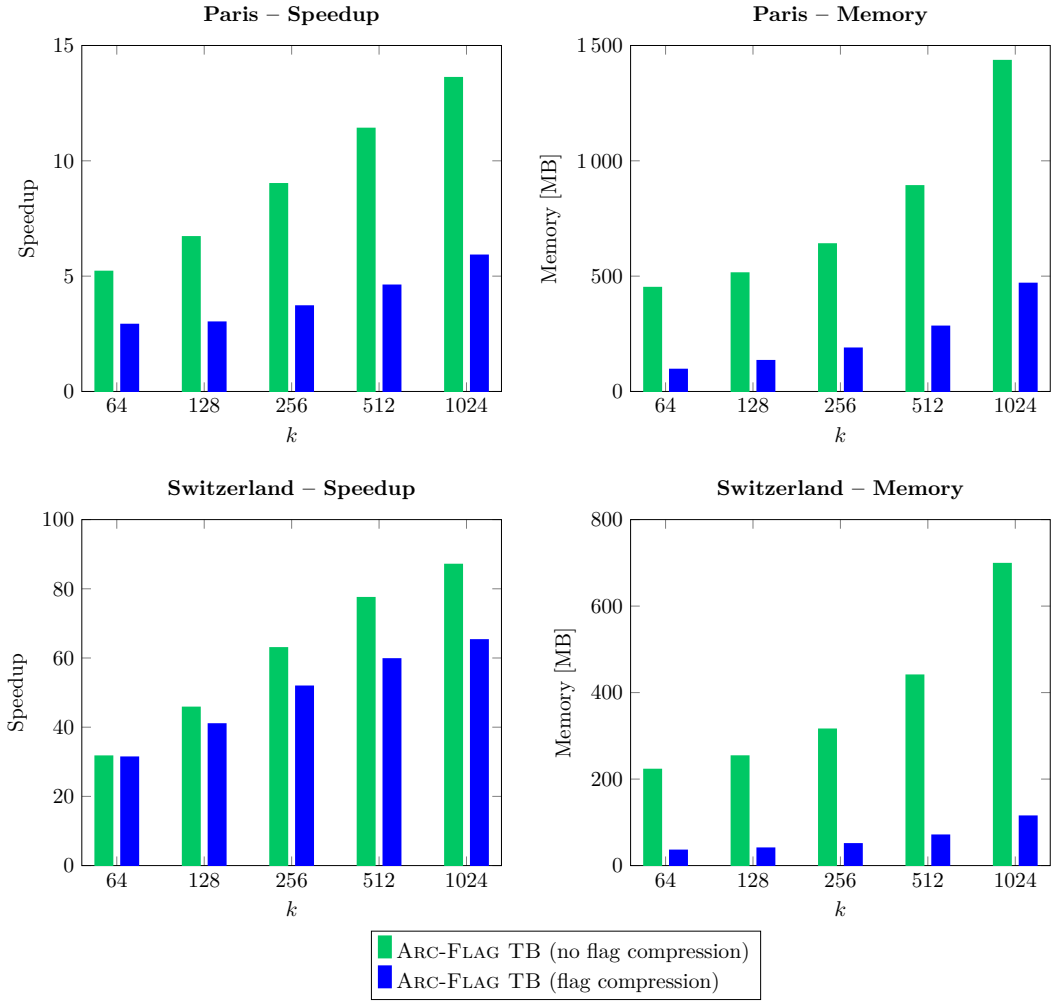

\subparagraph*{\arcflagtb Performance.}
Performance measurements for \arcflagtb, including the impact of flag compression and the number of cells~$\numCells$, are shown in Table~\ref{tab:results}.
For each configuration, we performed \numprint{10000} queries with the source and target stops chosen uniformly at random, and the departure time chosen uniformly at random within the first day of the timetable.
As expected, the preprocessing time is mostly unaffected by~$\numCells$.
The fastest configuration of \arcflagtb is with~$\numCells=\numprint{1024}$ and without flag compression.
Compared to \tb, it achieves a speedup of 143.2 on Germany, 13.6 on Paris, 86.2 on Sweden and 87.0 on Switzerland.
Even without compression, the memory consumption for the computed flags is moderate at roughly~1\,GB for the smaller networks and~5\,GB for Germany.
On all networks except Paris, flag compression is very effective: it reduces the memory consumption by a factor of $6$--$8$ at the expense of $20$--$40\%$ of additional query time.
On Paris, the compression is less successful but still reduces the memory consumption by a factor of~$3$ while roughly doubling the query time.
Figure~\ref{fig:resultplot} plots the speedup over \tb and the memory consumption, with and without flag compression, depending on~$\numCells$.
While the performance gains from doubling the number of cells eventually decline, they still remain strong up to~$\numCells=\numprint{1024}$.
The rate of incorrectly answered queries is around~$5\%$ on the country networks and~$1\%$ on Paris, and only slightly increases with~$\numCells$.
The differences in the results between Paris and the other networks are explained by the fact that Paris is a dense metropolitan network and therefore has a less hierarchical structure.
Similar discrepancies in the performance between metropolitan networks and country networks were observed for \algoname{Transfer Patterns}~\cite{Bas10} and \algoname{TB-CST}~\cite{Wit16}.

\subparagraph*{Result Quality.}
Table~\ref{tab:fixes} shows the impact of departure time buffering and flag augmentation on the result quality of \arcflagtb.
Departure time buffering significantly increases the preprocessing time, but this pays off in terms of the error rate, which is reduced from almost~$30\%$ to~$6\%$.
Flag augmentation on its own also reduces the error rate, but not as much.
Combining both only slightly reduces the error rate compared to departure time buffering alone, which indicates that the scenario depicted in Figure~\ref{fig:flag-augmentation} is rare.
We also tested a configuration in which most pruning rules of the \tb transfer generation were disabled.
Nearly all queries were answered correctly in this configuration, although this approximately doubled the preprocessing time and query time.

\begin{table}[t!]
    \centering
    \caption{Impact of departure time buffering (Buf.), flag augmentation (Aug.) and \tb preprocessing pruning rules (\tb) on the performance and success rate of \arcflagtb, measured on the Switzerland network.
        Query times are measured without flag compression.}
    \label{tab:fixes}
    \begin{tabular}{@{\,}cccrrrr@{\,}}
        \toprule
        \multirow{2}{*}{Buf.} & \multirow{2}{*}{Aug.} & \multirow{2}{*}{\tb} & \multirow{2}{*}{\thead{Prepro.\\ $\left[ \mathrm{hh}{:}\mathrm{mm}{:}\mathrm{ss}\right]$}} & \multirow{2}{*}{\thead{Query time \\ $\left[ \mu s\right]$}} & \multicolumn{2}{c}{Success rate $[\%]$} \\
        \cmidrule(){6-7}
        & & & & & Queries & Journeys \\
        \midrule
        $\circ$ & $\circ$ & $\bullet$ & \printTime{0}{5}{03} & \numprint{46} & 70.37 & 75.54 \\
        $\circ$ & $\bullet$ & $\bullet$ & \printTime{0}{5}{05} & \numprint{56} & 81.87 & 85.39 \\
        $\bullet$ & $\circ$ & $\bullet$ & \printTime{0}{13}{13} & \numprint{77} & 94.05 & 95.70 \\
        $\bullet$ & $\bullet$ & $\bullet$ & \printTime{0}{13}{14} & \numprint{81} & 94.80 & 96.29 \\
        $\bullet$ & $\bullet$ & $\circ$ & \printTime{0}{29}{41} & \numprint{184} & 99.50 & 99.70 \\
        \bottomrule
    \end{tabular}
\end{table}

\subparagraph*{TB-CST.}
Finally, we compare \arcflagtb against Witt's implementation of \algoname{TB-CST} with split trees~\cite{Wit16} on our networks.
The results are shown in Table~\ref{tab:tbcst-results}.
We do not report the performance of \algoname{TB-CST} with unsplit prefix trees since the precomputed data requires over~100\,GB of memory even on the smaller networks.
Excluding the partitioning step, which always took~10 minutes in our experiments, the precomputation time of \arcflagtb is~$2$--$6$ times higher, depending on the network.
Although both techniques perform a one-to-all Profile-\tb search from every stop, our algorithm additionally performs departure time buffering, which increases the precomputation time.
The remaining difference, which amounts to a factor of~2 on the Switzerland network, is due to the fact that our implementation of Profile-\tb is less optimized than Witt's.
The memory consumption of \arcflagtb is much lower than that of \algoname{TB-CST}, even with~\numprint{1024} cells and without flag compression.
Query times are similar or better on all networks except Germany.
Note that the cells on the Germany network are especially larger because the network is by far the largest.
We expect that \arcflagtb could achieve much lower query times here if more cells were used.

Overall, \arcflagtb matches the query performance of \algoname{TB-CST} while requiring much less space.
This is for two reasons:
Firstly, the query time of \algoname{TB-CST} is dominated by the time required to construct the query graph.
\arcflagtb does not require this step.
Secondly, the \algoname{TB-CST} query algorithm must reconstruct the earliest reachable trip of each used line at query time, whereas \arcflagtb can rely directly on the precomputed transfers.
Furthermore, we observe that \algoname{TB-CST} has a much higher error rate on Sweden and Switzerland than \arcflagtb.
We expect that this is because \arcflagtb aggregates the flags by cell.
Thus, even if the precomputation fails to find a required journey to a particular target stop, the transfers in that journey may still be flagged if they occur in journeys to other target stops in the same cell.

\begin{table}[htbp]
    \centering
    \caption{Performance of \algoname{TB-CST} with split trees for \numprint{10000} random queries.
        Note that we were not able to run~\algoname{TB-CST} queries on the Germany network due to issues with the provided code.
        We instead list the query time reported in~\cite{Wit16}.}
    \label{tab:tbcst-results}
    \begin{tabular}{@{\,}lrrrrrrr@{\,}}
        \toprule
        \multirow{2}{*}{Network} & \multirow{2}{*}{\thead{Prepro.\\ $\left[ \mathrm{hh}{:}\mathrm{mm}{:}\mathrm{ss}\right]$}} & \multirow{2}{*}{\thead{Query time \\ $\left[ \mu s\right]$}} & \multirow{2}{*}{\thead{Memory \\ $\left[ \mathrm{MB} \right]$}} & \multicolumn{2}{c}{Success rate $[\%]$} & \\
        \cmidrule(){5-6}
        & & & & Queries & Journeys \\
        \midrule
        Germany & \printTime{6}{36}{27} & (\numprint{156}) & \numprint{114080} & -- & -- \\
        Paris & \printTime{0}{20}{30} & \numprint{507} & \numprint{6992} & 98.98 & 99.05 \\
        Sweden & \printTime{0}{7}{42} & \numprint{91} & \numprint{3400} & 75.99 & 91.67 \\
        Switzerland & \printTime{0}{2}{22} & \numprint{66} & \numprint{1586} & 80.72 & 89.88  \\
        \bottomrule
    \end{tabular}
\end{table}

\section{Conclusion}
\label{sec:conclusion}
We developed \arcflagtb, a speedup technique for public transit journey planning which combines \algoname{Arc-Flags} and \algoname{Trip-Based Public Transit Routing}~(\tb).
We demonstrated that the stronger pruning rules of \tb allow our approach to overcome previous obstacles in applying \algoname{Arc-Flags} to public transit networks.
This allows \arcflagtb to achieve up to two orders of magnitude speedup over \tb.
Compared to \algoname{TB-CST}, a state-of-the-art speedup technique for \tb, our algorithm achieves roughly the same query times with a similar precomputation time and only a fraction of the memory consumption.
Unlike \algoname{TB-CST}, the query performance and memory consumption are configurable via the number of cells in the computed network partition.
Currently, both algorithms answer some queries incorrectly due to an issue with the \tb precomputation phase.
However, we showed that the error rate of \arcflagtb is low and discussed approaches for resolving the issue.
In the future, it would be interesting to examine whether the performance of \arcflagtb can still be achieved with a subquadratic precomputation phase which only runs searches from the boundary nodes of the partition.

\bibliography{bibliography.bib}

\begin{thebibliography}{10}

\bibitem{Bas09}
Hannah Bast.
\newblock {Car or Public Transport -- Two Worlds}.
\newblock In {\em Efficient Algorithms}, volume 5760 of {\em Lecture Notes in
  Computer Science~(LNCS)}, pages 355--367. Springer, 2009.
\newblock \href {https://doi.org/10.1007/978-3-642-03456-5_24}
  {\path{doi:10.1007/978-3-642-03456-5_24}}.

\bibitem{Bas10}
Hannah Bast, Erik Carlsson, Arno Eigenwillig, Robert Geisberger, Chris
  Harrelson, Veselin Raychev, and Fabien Viger.
\newblock {Fast Routing in Very Large Public Transportation Networks using
  Transfer Patterns}.
\newblock In {\em Proceedings of the~18th Annual European Symposium on
  Algorithms~(ESA'10)}, volume 6346 of {\em Lecture Notes in Computer
  Science~(LNCS)}, pages 290--301. Springer, 2010.
\newblock \href {https://doi.org/10.1007/978-3-642-15775-2_25}
  {\path{doi:10.1007/978-3-642-15775-2_25}}.

\bibitem{Bas16b}
Hannah Bast, Daniel Delling, Andrew Goldberg, Matthias Müller-Hannemann,
  Thomas Pajor, Peter Sanders, Dorothea Wagner, and Renato~F. Werneck.
\newblock {Route Planning in Transportation Networks}.
\newblock In {\em Algorithm Engineering: Selected Results and Surveys}, volume
  9220 of {\em Lecture Notes in Computer Science~(LNCS)}, pages 19--80.
  Springer, 2016.
\newblock \href {https://doi.org/10.1007/978-3-319-49487-6_2}
  {\path{doi:10.1007/978-3-319-49487-6_2}}.

\bibitem{Bas16}
Hannah Bast, Matthias Hertel, and Sabine Storandt.
\newblock {Scalable Transfer Patterns}.
\newblock In {\em Proc.~18th Workshop on Algorithm Engineering and
  Experiments~(ALENEX'16)}, pages 15--29. Society for Industrial and Applied
  Mathematics~(SIAM), 2016.
\newblock \href {https://doi.org/10.1137/1.9781611974317.2}
  {\path{doi:10.1137/1.9781611974317.2}}.

\bibitem{Bau09}
Reinhard Bauer and Daniel Delling.
\newblock {SHARC: Fast and Robust Unidirectional Routing}.
\newblock {\em Journal of Experimental Algorithmics~(JEA)}, 14:4.1--4.29, 2009.
\newblock \href {https://doi.org/10.1145/1498698.1537599}
  {\path{doi:10.1145/1498698.1537599}}.

\bibitem{Bau11}
Reinhard Bauer, Daniel Delling, and Dorothea Wagner.
\newblock {Experimental Study of Speed Up Techniques for Timetable Information
  Systems}.
\newblock {\em Networks}, 57:38--52, 2011.
\newblock \href {https://doi.org/10.1002/net.20382}
  {\path{doi:10.1002/net.20382}}.

\bibitem{Ber09}
Annabell Berger, Daniel Delling, Andreas Gebhardt, and Matthias
  Müller-Hannemann.
\newblock {Accelerating Time-Dependent Multi-Criteria Timetable Information is
  Harder Than Expected}.
\newblock In {\em Proceedings of the~9th Workshop on Algorithmic Approaches for
  Transportation Modelling, Optimization, and Systems~(ATMOS'09)}, volume~12 of
  {\em OpenAccess Series in Informatics~(OASIcs)}, pages 2:1--2:21. Schloss
  Dagstuhl -- Leibniz-Zentrum für Informatik, 2009.
\newblock \href {https://doi.org/10.4230/OASIcs.ATMOS.2009.2148}
  {\path{doi:10.4230/OASIcs.ATMOS.2009.2148}}.

\bibitem{Bro04}
Gerth~S. Brodal and Riko Jacob.
\newblock {Time-dependent Networks as Models to Achieve Fast Exact Time-table
  Queries}.
\newblock In {\em Proceedings of the~3rd Workshop on Algorithmic Approaches for
  Transportation Modelling, Optimization, and Systems~(ATMOS'03)}, volume~92,
  pages 3--15. Elsevier, 2004.
\newblock \href {https://doi.org/10.1016/j.entcs.2003.12.019}
  {\path{doi:10.1016/j.entcs.2003.12.019}}.

\bibitem{Coh03}
Edith Cohen, Eran Halperin, Haim Kaplan, and Uri Zwick.
\newblock {Reachability and Distance Queries via 2-Hop Labels}.
\newblock {\em SIAM Journal on Computing~(SICOMP)}, 32(5):1338--1355, 2003.
\newblock \href {https://doi.org/10.1137/S0097539702403098}
  {\path{doi:10.1137/S0097539702403098}}.

\bibitem{Del15}
Daniel Delling, Julian Dibbelt, Thomas Pajor, and Renato~F. Werneck.
\newblock {Public Transit Labeling}.
\newblock In {\em Proceedings of the~14th International Symposium on
  Experimental Algorithms~(SEA'15)}, volume 9125 of {\em Lecture Notes in
  Computer Science~(LNCS)}, pages 273--285. Springer, 2015.
\newblock \href {https://doi.org/10.1007/978-3-319-20086-6_21}
  {\path{doi:10.1007/978-3-319-20086-6_21}}.

\bibitem{Del09c}
Daniel Delling, Thomas Pajor, and Dorothea Wagner.
\newblock {Engineering Time-Expanded Graphs for Faster Timetable Information}.
\newblock In {\em Robust and Online Large-Scale Optimization: Models and
  Techniques for Transportation Systems}, volume 5868 of {\em Lecture Notes in
  Computer Science~(LNCS)}, pages 182--206. Springer, 2009.
\newblock \href {https://doi.org/10.1007/978-3-642-05465-5_7}
  {\path{doi:10.1007/978-3-642-05465-5_7}}.

\bibitem{Del15b}
Daniel Delling, Thomas Pajor, and Renato~F. Werneck.
\newblock {Round-Based Public Transit Routing}.
\newblock {\em Transportation Science}, 49:591--604, 2015.
\newblock \href {https://doi.org/10.1287/trsc.2014.0534}
  {\path{doi:10.1287/trsc.2014.0534}}.

\bibitem{Dij59}
Edsger~W. Dijkstra.
\newblock {A Note on Two Problems in Connexion with Graphs}.
\newblock {\em Numerische Mathematik}, 1:269--271, 1959.
\newblock \href {https://doi.org/10.1007/BF01386390}
  {\path{doi:10.1007/BF01386390}}.

\bibitem{Dis08}
Yann Disser, Matthias Müller-Hannemann, and Mathias Schnee.
\newblock {Multi-criteria Shortest Paths in Time-Dependent Train Networks}.
\newblock In {\em Proceedings of the~7th International Workshop on Experimental
  and Efficient Algorithms~(WEA'08)}, volume 5038 of {\em Lecture Notes in
  Computer Science~(LNCS)}, pages 347--361. Springer, 2008.
\newblock \href {https://doi.org/10.1007/978-3-540-68552-4_26}
  {\path{doi:10.1007/978-3-540-68552-4_26}}.

\bibitem{Han80}
Pierre Hansen.
\newblock {Bicriterion Path Problems}.
\newblock In {\em {Multiple Criteria Decision Making Theory and Application}},
  volume 177 of {\em Lecture Notes in Economics and Mathematical Systems},
  pages 109--127. Springer, 1980.
\newblock \href {https://doi.org/10.1007/978-3-642-48782-8_9}
  {\path{doi:10.1007/978-3-642-48782-8_9}}.

\bibitem{Hil09}
Moritz Hilger, Ekkehard Köhler, Rolf~H. Möhring, and Heiko Schilling.
\newblock {Fast Point-to-Point Shortest Path Computations with Arc-Flags}.
\newblock In {\em The Shortest Path Problem: Ninth DIMACS Implementation
  Challenge}, volume~74 of {\em DIMACS Series in Discrete Mathematics and
  Theoretical Computer Science}, pages 41--72. American Mathematical
  Society~(AMS), 2009.
\newblock \href {https://doi.org/10.1090/dimacs/074/03}
  {\path{doi:10.1090/dimacs/074/03}}.

\bibitem{Moe06}
Rolf~H. Möhring, Heiko Schilling, Birk Schütz, Dorothea Wagner, and Thomas
  Willhalm.
\newblock {Partitioning Graphs to Speed Up Dijkstra's Algorithm}.
\newblock {\em Journal of Experimental Algorithmics~(JEA)}, 11:2.8:1--2.8:29,
  2006.
\newblock \href {https://doi.org/10.1007/11427186_18}
  {\path{doi:10.1007/11427186_18}}.

\bibitem{Mue07b}
Matthias Müller-Hannemann and Mathias Schnee.
\newblock {Finding All Attractive Train Connections by Multi-Criteria Pareto
  Search}.
\newblock In {\em Algorithmic Methods for Railway Optimization}, volume 4359 of
  {\em Lecture Notes in Computer Science~(LNCS)}, pages 246--263. Springer,
  2007.
\newblock \href {https://doi.org/10.1007/978-3-540-74247-0_13}
  {\path{doi:10.1007/978-3-540-74247-0_13}}.

\bibitem{Pyr08}
Evangelia Pyrga, Frank Schulz, Dorothea Wagner, and Christos~D. Zaroliagis.
\newblock {Efficient Models for Timetable Information in Public Transportation
  Systems}.
\newblock {\em Journal of Experimental Algorithmics~(JEA)}, 12:2.4:1--2.4:39,
  2008.
\newblock \href {https://doi.org/10.1145/1227161.1227166}
  {\path{doi:10.1145/1227161.1227166}}.

\bibitem{San13}
Peter Sanders and Christian Schulz.
\newblock {Think Locally, Act Globally: Highly Balanced Graph Partitioning}.
\newblock In {\em Proceedings of the~12th International Symposium on
  Experimental Algorithms~(SEA'13)}, volume 7933 of {\em Lecture Notes in
  Computer Science~(LNCS)}, pages 164--175. Springer, 2013.
\newblock \href {https://doi.org/10.1007/978-3-642-38527-8_16}
  {\path{doi:10.1007/978-3-642-38527-8_16}}.

\bibitem{Wit15}
Sascha Witt.
\newblock {Trip-Based Public Transit Routing}.
\newblock In {\em Proceedings of the~23rd Annual European Symposium on
  Algorithms~(ESA'15)}, volume 9294 of {\em Lecture Notes in Computer
  Science~(LNCS)}, pages 1025--1036. Springer, 2015.
\newblock \href {https://doi.org/10.1007/978-3-662-48350-3_85}
  {\path{doi:10.1007/978-3-662-48350-3_85}}.

\bibitem{Wit16}
Sascha Witt.
\newblock {Trip-Based Public Transit Routing Using Condensed Search Trees}.
\newblock In {\em Proc.~16th Workshop on Algorithmic Approaches for
  Transportation Modelling, Optimization, and Systems~(ATMOS'16)}, volume~54 of
  {\em OpenAccess Series in Informatics~(OASIcs)}, pages 10:1--10:12. Schloss
  Dagstuhl -- Leibniz-Zentrum für Informatik, 2016.

\end{thebibliography}

\end{document}